\documentclass[]{JFM-FLM_Au}
\hypersetup{
	citecolor = blue
}
\usepackage{subcaption}
\usepackage{amssymb}
\usepackage[normalem]{ulem}

\newcommand{\bU}{\boldsymbol{U}}
\newcommand{\bex}{\hat{\mathbf{e}}_x}
\newcommand{\bey}{\hat{\mathbf{e}}_y}
\newcommand{\bu}{\boldsymbol{u}}
\newcommand{\bn}{\hat{\mathbf{n}}}
\newcommand{\bx}{\boldsymbol{x}}
\newcommand{\bD}{\boldsymbol{D}}
\newcommand{\tnabla}{\tilde{\nabla}}
\newcommand{\tPsi}{\tilde{\Psi}}
\newcommand{\tH}{\tilde{H}}
\newcommand{\tx}{\tilde{x}}
\newcommand{\ty}{\tilde{y}}
\newcommand{\tOmega}{\tilde{\Omega}}
\newcommand{\tT}{\tilde{T}}
\newcommand{\tU}{\tilde{U}}
\newcommand{\tc}{\tilde{c}}

\lefttitle{S.N. Jadhav, S. Ravichandran and R. Dasgupta}
\righttitle{Journal of Fluid Mechanics}

\title{Oceanic surface waves from wind-drift-layer instability: exponential growth to finite-amplitude states}

\author{Sayali N. Jadhav\aff{1}\corresp{\email{sayalijadhav2016@gmail.com}}, S. Ravichandran\aff{2} \and Ratul Dasgupta\aff{1}\corresp{\email{dasgupta.ratul@gmail.com}}}

\affiliation{\aff{1}Department of Chemical Engineering, Indian Institute of Technology Bombay, Mumbai, India 400 076
\aff{2}Centre for Climate Studies, Indian Institute of Technology Bombay, Mumbai, India 400 076}

\begin{document}
\maketitle
\begin{abstract}
Oceanic wind-wave genesis via shear-induced linear instability and the propagation and breaking of finite-amplitude waves of permanent form have traditionally been studied as decoupled problems. While experimental evidence indicates connection between the two regimes, a first principles demonstration that the latter can emerge from the former remains challenging. Here, we bridge this gap by investigating gravity-dominated surface waves arising from an inviscid, shear-driven Holmboe instability. The background state comprises a uniform air current and a constant-vorticity water drift layer of finite thickness overlying deep irrotational water. In the high-Bond-number and low-Froude-number regime, this configuration supports Holmboe-unstable modes beyond a critical Froude number. Two-phase simulations of the incompressible Navier–Stokes equations, initialised with the fastest-growing eigenmode, exhibit initial exponential growth in excellent agreement with linear theory. In the subsequent nonlinear stage preceding breaking, finite-amplitude propagating waveforms develop. Spatio-temporal spectral analysis reveals bound harmonics locked to the primary carrier wave (the fastest-growing mode), coexisting with residual free-wave components. This indicates that an infinitesimal, linearly unstable mode can evolve into a nonlinear propagating waveform of approximately permanent form under background shear. To characterise these waveforms theoretically, we formulate an exact nonlinear boundary-value problem for steady-state propagating waves including background shear. The resulting wave profiles show very good agreement with time-averaged profiles extracted from the transient simulations in the nonlinear stage. These findings extend the theory of \cite{milinazzo1990effect} and complement the wind-wave observations of \cite{lake1978new}, demonstrating that Stokes-like, sheared propagating surface gravity waves of finite-amplitude, can naturally emerge during the nonlinear evolution of a wind-drift-induced linear instability.
\end{abstract}

\section{Introduction}\label{sec:Introduction}

The generation mechanism(s) of surface waves in the open ocean, directly from wind or due to currents in the upper ocean, and their subsequent spatio-temporal fate are topics of longstanding research interest  \citep{miles1957generation,longuet1997progress,mitsuyasu2002historical, pizzo2021does}. Waves in the ocean are typically of moderate steepness (maximum slope), observed in the range $0.05-0.2$ \citep{longuet1994crest}. This precludes ocean waves from reaching the (strongly nonlinear) limiting form of Stokes waves \citep{deconinck2023dominant}. Even at these moderate steepnesses, however, ocean waves do break, thereby seeding multi-phase multi-scale processes, which culminate in white-caps and sea-spray \citep{andreas1992sea,andreas1995spray,veron2015ocean,richter2016ocean}. While fig. 1 in \cite{pierson1955wind} captures the already formidable theoretical foundations of early wind-wave science, our understanding of the subject - encompassing wave generation, breaking, and their measurement and quantification - has since expanded exponentially, see table $\#1$ in \cite{mitsuyasu2002historical} and \cite{pizzo2021does,deike2022mass}.
	
Given the vast range of length and time-scales of some of these aforementioned processes, a first-principles spatio-temporal description of wind-wave generation  - commencing from a flat interface and extending through wave-breaking and beyond, with accompanying mechanistic insight - remains an exceptionally challenging open problem. Several studies have thus confined attention towards understanding specific subsets of the whole, often in a decoupled manner. For instance, two well-studied wave-generating mechanisms are background shear induced, exponentially growing waves \citep{miles1957generation,morland1993effect,young2014generation} and waves excited by turbulent air pressure fluctuations advected by the shear \citep{phillips1957generation}. Separately, a substantial body of computational and experimental work has investigated the breaking of finite-amplitude propagating waves, albeit without directly accounting for background wind shear \citep{chen1999two,DeaneStokes2002,deike2015capillary,deike2016air,chan2021turbulent}. In addition to seeding the linear instability, background shear (in air or water) can significantly distort the form and phase speed of the finite-amplitude travelling waveforms, potentially resulting from the linear instability. The quasi-linear theory of \cite{janssen1982quasilinear}, for example, investigated the emergence of steady-state waveforms via the quenching of the linear instability, as the background shear (in air) gets distorted at large times, also predicting the form of the wave energy spectrum (see fig. $2$ in \cite{janssen1982quasilinear}), under the parallel flow approximation. 
	
The energy spectrum of wind-waves has been of specific interest, including mathematical models for its spatio-temporal evolution \citep{Hasselmann1960,Barnett1968}. Here, a series of experimental studies in the late 1970s \citep{ramamonjiarisoa1979modulation,lake1978new} discovered an important role of nonlinearity on the wind-wave energy spectrum. At frequencies close to twice the peak of this spectrum, the phase speeds of the Fourier components were observed to be closely matching that of the peak spectral frequency. This was due to the generation of nonlinear, bound wave components propagating at the same speed as the peak spectral (free wave) component, please see fig. $2$ in \cite{mitsuyasu1979dispersion} and fig. $8$ in \cite{mitsuyasu2015reminiscences}. Now, it has been well-known that the finite-amplitude Stokes wave in a shear-free background, frequently used to initialise breaking wave simulations \citep{chen1999two,deike2015capillary,chan2021turbulent}, comprises a single free wave and an infinite sequence of bound harmonics travelling at the same phase speed, thus preserving the wave form. Indeed, \cite{lake1978new, yuen1979new} proposed that the nonlinear wind-wave system could be described as such a wave-train to a first approximation. Naturally, one may ask how does background shear affect and alter these conclusions. For example, could a linear instability of the flat air-water interface - driven by background shear in water (such as a wind-drift-layer or an upper-ocean current in the water) - result in a spatio-temporal waveform that, in a time-averaged sense, be well described in a first approximation as a shear-modified, Stokes-like progressive wave of permanent form? 

Here we demonstrate computationally and theoretically, that this is indeed so, with implications for our understanding of surface wave generation processes in the open ocean. With this motivation, sections \ref{subsec:LinInstab} and \ref{subsec:FiniteAmp} briefly describe the shear-induced linear instability for our choice of the background profile and permanent-form, finite-amplitude Stokes-like propagating waves modified by the background shear, respectively.

\subsection{Linear (in)stability of a flat air-water interface}\label{subsec:LinInstab}
As wind blows over an initially quiescent sea surface, viscosity-mediated momentum flux generates a turbulent shear layer on the air side \citep{belcher1993turbulent,garratt1994atmospheric,buckley2016structure} as well as a wind-drift (shear) layer in the upper sea \citep{phillips1974wave,wu1975wind, shrira1993surface}. The thickness of these respective layers evolve with prolonged wind forcing. This setup makes the air-water system susceptible to the classical linear instabilities of parallel shear flows with density stratification \citep{rayleigh1879stability,goldstein1931stability,Schmid2001stability}. The most prominent air-side surface wave-generation mechanism stems from the shear-flow linear instability of \citet{miles1957generation} (predated by the sheltering hypothesis of \citealp{jeffreys1925formation}), which crucially relies on non-zero curvature of the background shear profile at the (critical) location where the perturbation wave speed matches that of the background shear. This linear instability mechanism complements the turbulent pressure-fluctuation mechanism of \citet{phillips1957generation}, the latter yielding wave growth  algebraic in time. The Miles critical-layer model in particular, has been investigated heavily \citep{valenzuela1976growth, plant1982relationship, peirson2008wind} with analytical results continuing to the present \citep{bonfils2022asymptotic,seitz2026role}. Various approximate background velocity profiles have been considered including quasi-laminar logarithmic profiles \citep{miles1957generation} and error-function and other similar profiles  \citep{lock1954hydrodynamic, morland1993effect,kumar2026waves}. While qualitative match with theoretical predictions have indeed been observed \citep{hristov2003dynamical}, quantitative mismatch persists; see discussion in page $5$, second paragraph in \cite{sullivan2010dynamics}. These observations have also motivated recent theoretical investigations on spatial growth of wind-driven waves \citep{kumar2024spatial}, as opposed to temporal analysis. It is important to recall that while these aforementioned background shear profiles are fundamentally generated by viscous diffusion, their linear stability can often be understood, in a first approximation, within an inviscid framework. This is justified because the advective timescale of the perturbed flow ($t_c \sim l_c/u_c$, where $u_c$ and $l_c$ are the characteristic velocity and length scales, respectively) is vastly smaller than the viscous diffusion timescale ($t_\nu \sim l_c^2/\nu_c$, where $\nu_c$ is the characteristic kinematic viscosity). At the oceanic scale, $l_c$ is of the order of a gravity wavelength (typically $1$ m), $u_c \sim \mathbb{O}(1)$ m/s being the natural phase-speed of this wave. This leads to an asymptotically large Reynolds number ($Re \equiv u_c l_c/\nu_c \sim \mathbb{O}(10^{6})$), allowing the perturbation dynamics to be dominated by linearised inertia at leading order. 	 

An early three-dimensional direct numerical simulation (DNS) of wind-wave generation from an initially flat air-water interface - using linearised interfacial boundary conditions and supporting both Phillips and Miles-type wave-growth mechanisms - was carried out by \cite{lin2008direct} (see also \citealt{li2022principal} for more recent work). More recently, \cite{wu2021wind} investigated two-dimensional DNS of a gravity-capillary wave - initialised as a small-amplitude Fourier mode with linear orbital velocities in the water - and forced by a linear background shear profile in the air. Alongside the concurrent development of a viscous subsurface drift layer, the authors reported exponential wave growth that progressed into distinct nonlinear regimes depending on the Bond number, recovering parasitic capillary ripples, and spilling breakers. Subsequently, \cite{wu2022revisiting} reported three-dimensional DNS of wave growth underneath a pressure-gradient-driven turbulent airflow. Initialising the interface with a small-amplitude unsheared Stokes wave, forced subsequently by the turbulent air, their simulations described the progression from exponential growth to nonlinear steepening and the onset of wave breaking. While a viscous subsurface drift layer also developed simultaneously in the water, its influence on wave growth was found to be secondary for the parameter regime investigated; also see \cite{scapin2025momentum,scapin2026growth} for related studies as well as \cite{yang2013dynamic,hao2019wind} for (large-eddy simulation) LES work. We also refer the reader to section $1.1$ in \cite{kumar2026waves} where a few other simulational studies with background shear are discussed. 

\begin{figure}
    \centering
    \begin{subfigure}[c]{0.54\textwidth}
        \centering
        \includegraphics[width=\textwidth]{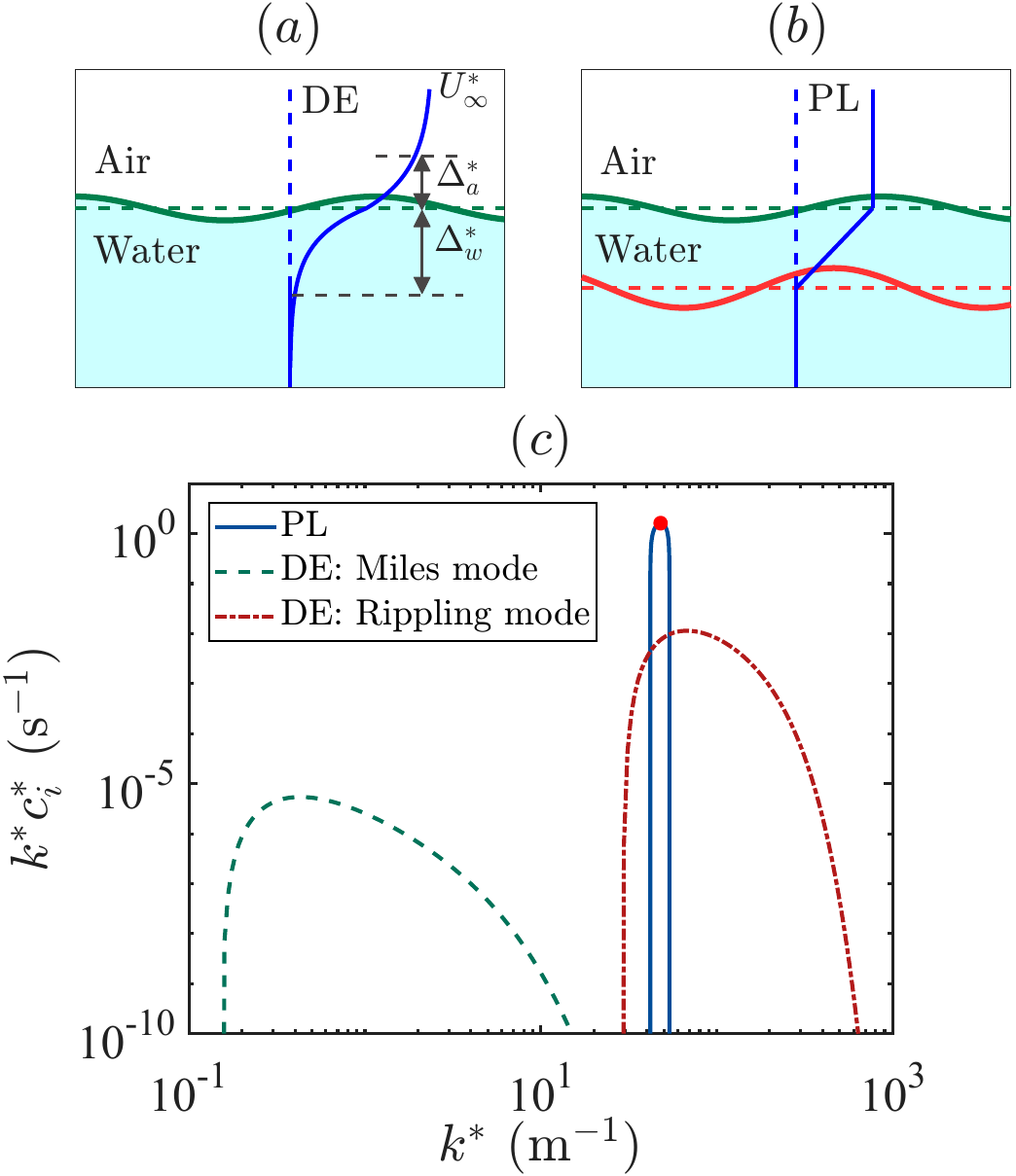} 
        \label{fig:miles_mode}
    \end{subfigure}
    \hfill 
    \begin{subfigure}[c]{0.42\textwidth}
        \centering
        \includegraphics[width=\textwidth]{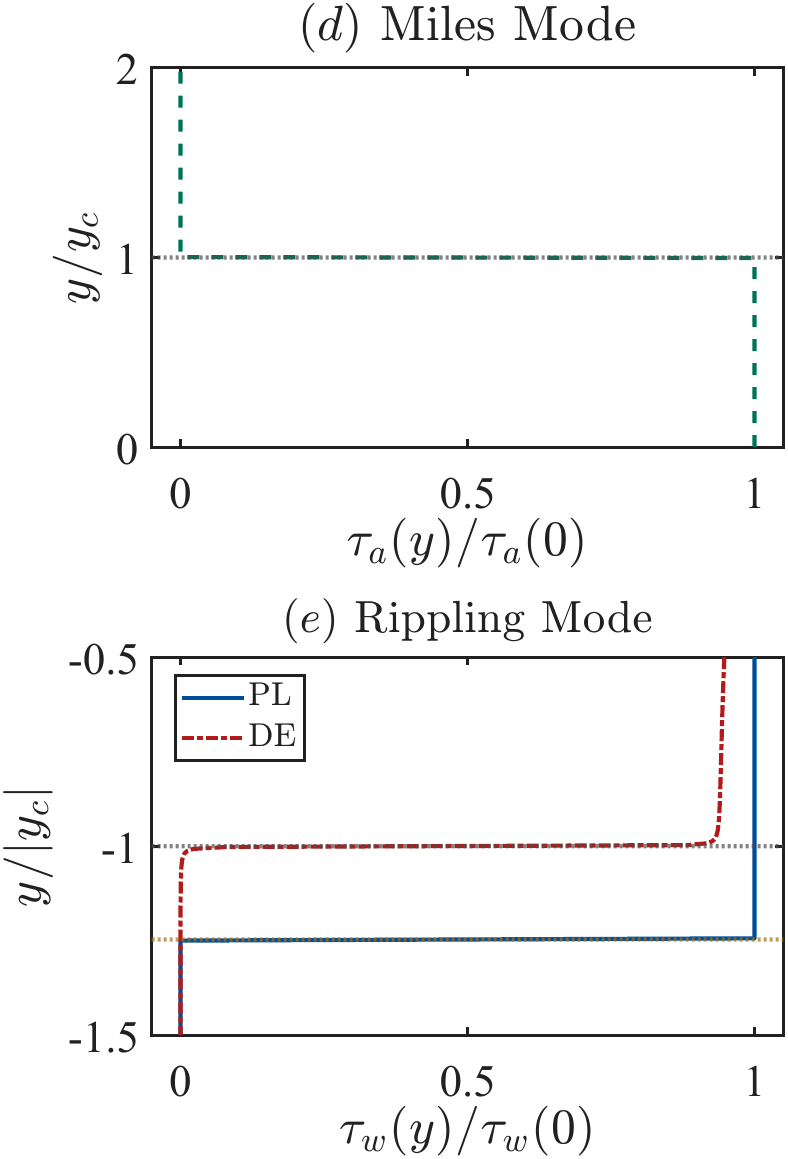} 
        \label{fig:rippling_mode}
    \end{subfigure}
    
    \caption{Comparison of linear instability mechanisms and growth rates for two background horizontal velocity profiles (blue curves in a, b):
    	(a) the double-exponential (DE) profile with continuously varying shear in both air and water \citep{young2014generation}, and
    	(b) the piecewise-linear (PL) profile used in this study, featuring uniform shear over a finite depth $\Delta^*_w$ in water and zero shear in air. The green curves depict the air-water surface waves, while the red curve in (b) illustrates the vorticity wave at the base of the water shear layer for the PL profile.
    	(c) Dimensional growth rate $k^*c^*_i$ (units of $\text{s}^{-1}$) versus wavenumber $k^*$ ($\text{m}^{-1}$). For the DE profile (evaluated using Eqn.~(2.15) in \citealt{young2014generation}) and the chosen parameter regime, the water-side ``rippling'' instability (red dash-dotted curve) is roughly three orders of magnitude stronger than the air-side ``Miles'' instability (green dashed curve). For the PL profile (blue solid curve, evaluated using eqn.~\ref{eq: dispersion_relation_LSA_nondimensional}), the water-side growth rate is another two orders of magnitude higher than the DE rippling mode, with the red dot marking the most unstable mode.
    	(d, e) Vertical variation of the (inviscid) Reynolds stress $\tau(y) = -\overline{\rho u v}$, normalized by its surface value $\tau(0)$ for the fastest growing mode(s), contrasting the underlying instability mechanisms:
    	(d) the air-side Miles mode ($\tau_a$, green dashed curve) for the DE profile, and
    	(e) the water-side rippling mode (red dash-dotted curve) comparing the Reynolds stress from the DE profile and the PL profile (blue solid curve). For the DE profile, sharp variations in Reynolds stress occur at the air and water critical layers, $y/|y_c| = \pm 1$ ($y_c$ being the location where the wave phase speed matches the background flow speed, marked by horizontal gray dotted lines in (d),(e)). Whereas for the PL profile this abrupt variation occurs at the lower boundary of the water shear layer ($y = -\Delta^*_w$, located at $y/|y_c| \approx -1.247$, marked by dotted yellow line in (e)).
    	Results are shown for a surface drift velocity $U^*(y=0) = U^*_s = 80\text{ cm s}^{-1}$ and water shear-layer depth $\Delta^*_w = 5\text{ cm}$. For both DE and PL profiles, these correspond to the dimensionless parameters (formally defined in \S\ref{sec:Problem_statement}): Froude number $Fr \equiv \dfrac{U_s^*}{\sqrt{g^*\Delta_w^*}} = 1.143$, Bond number $Bo \equiv \dfrac{\rho_w^*g^*\Delta_w^{*2}}{T^*} = 340.277$, air-to-water density ratio $r = 0.001$. For the DE profile, the shear-layer depth ratio $\Delta^*_w/\Delta^*_a = 0.008$, and surface-to-freestream velocity ratio $U^*_s/U^*_{\infty} = 0.1$, see table \ref{tab:Tab3_nondim_numbers_LSA} for meaning of symbols.}
    \label{fig1:Comparison_with_DE}
\end{figure}

The recent linear stability analysis  of \cite{young2014generation} has crucially highlighted that the drift-layer induced instability (i.e. from background shear in water) can have maximal growth rates considerably larger than that of the air-side instability (from background shear in air). These authors studied the double exponential (DE hereafter) profile illustrated in Fig. \ref{fig1:Comparison_with_DE}a. Interestingly, this ``rippling-mode'' instability operates on the same physical mechanism as \cite{miles1957generation}, except that the critical layer is on the water side \citep{bonfils2023flow}.
As shown in Fig. \ref{fig1:Comparison_with_DE}c (left panel) (growth-rate versus wavenumber), the water-side shear induced instability can indeed dominate over the air-side instability in specific parameter regimes.
We are interested here in the parameter regime where an eventual nonlinear state arises from the linear instability induced by shear on the water side. A general analytical solution to the fully nonlinear problem with a curved background shear profile, such as the DE profile in fig. \ref{fig1:Comparison_with_DE}(a), is generally unavailable. To render both the linear instability and the subsequent finite-amplitude propagating wave solutions analytically accessible, idealized piecewise-linear (PL hereafter) profiles have historically been adopted as approximations to smooth background profiles e.g. see \cite{esch1962stability, stern1973capillary, morland1991waves,lawrence1991stability,longuet1998instabilities,baumstein1997nonlinear,carpenter2011instability}. 

In this study, we employ a similar PL profile having shear on the water side only, as illustrated in Fig. \ref{fig1:Comparison_with_DE}b. The subsurface drift layer is represented by a finite zone of constant shear, whereas the background air flow is taken to be uniform to preclude the onset of any subdominant air-side instability. This background configuration thus introduces two distinct interfaces: a density-vorticity interface at the air-water boundary and a vorticity interface at the velocity kink at the lower boundary of the drift layer. For the PL profile excluding the air side velocity, \citet{morland1991waves} reported growth rates an order of magnitude larger than those of equivalent smooth profiles. Consistent with this, Fig. \ref{fig1:Comparison_with_DE}c shows that the PL profile including air exhibits growth rates approximately two orders of magnitude larger than the water-side rippling mode of the equivalent DE profile. The fundamental distinction between the DE and PL instability mechanisms is reflected in the vertical variation of the inviscid wave-induced Reynolds stress, $\tau(y)=-\overline{\rho u v}$ (overbar indicates horizontal average over one wavelength), shown normalized by $\tau(0)$ in Fig. \ref{fig1:Comparison_with_DE}(d),(e). For the DE profile, continuous background profile curvature produces a sharp, localized stress variation across the critical layer ($y=y_c$) for both Miles and rippling modes. In contrast, the PL profile has constant vorticity within the subsurface shear layer, resulting in a constant Reynolds stress that drops abruptly to zero at the lower boundary of the shear-layer and remains zero in the underlying irrotational deep water. The relatively rapid instability growth makes the PL profile well suited for studying evolution of nonlinear propagating waves within finite time windows. These finite-amplitude propagating wave solutions with background shear (PL) are briefly introduced next.

\subsection{Finite amplitude propagating waves and the underlying linear instability}\label{subsec:FiniteAmp}

Finite-amplitude propagating waves of permanent form as obtained by solving the nonlinear, inviscid governing equations for various prescribed background velocity profiles, have been of fundamental interest independently. Some studies have accounted for air shear indirectly through modelling  pressure fluctuations therein and seeking finite amplitude propagating waves \citep{fedorov1998nonlinear,yewale2026windyseasurfacestokes}. Background profiles which have been studied include irrotational flow in the water, both in the absence of air \citep{chen1980numerical} and with a uniform air current \citep{saffman1982finite} or linear shear profiles in water \citep{simmen1985steady,dyachenko2019stokes}. Previously \cite{milinazzo1990effect} and more recently \citet{shelton2026nonlinear} 
reported finite-amplitude propagating wave solutions with piecewise-linear (PL) shear profiles, representing continuation of infinitesimal amplitude propagating modes, linearly stable under the sheared background state. Consequently, the finite-amplitude propagating waves studied by these authors cannot (directly) arise in the nonlinear stage of the primary linear instability, of interest here. 
\begin{figure}[ht!]
	\centering
	\includegraphics[width=\linewidth]{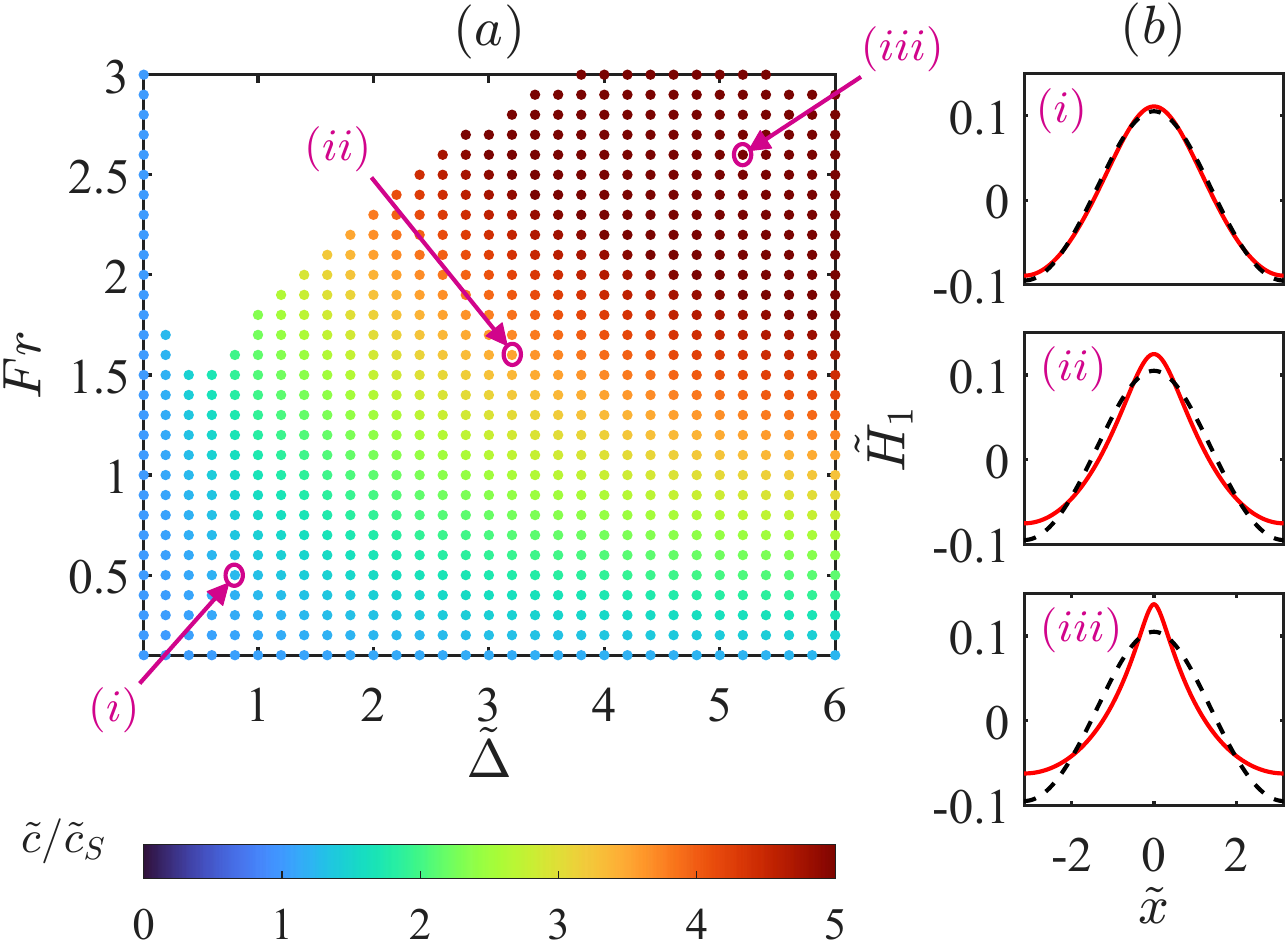}
	\caption{Each point on the plane of Froude number versus non-dimensional shear depth ($Fr-\tilde{\Delta}$) represents a finite-amplitude gravity-dominated wave with background shear. Here $Fr$ is the same as fig. \ref{fig1:Comparison_with_DE} and $\tilde{\Delta} \equiv k^*\Delta^*$, $k^*$ and $\Delta^*$ being the dimensional (primary) wavenumber of the nonlinear wave and the shear-depth in the PL profile (fig. \ref{fig1:Comparison_with_DE}(b)) respectively. These nonlinear waves represent continuation of infinitesimal amplitude, prograde Fourier modes (i.e. propagating in the same direction as the background surface drift velocity $U_s^*$ in a frame moving with $U_s^*$), linearly stable under the background PL profile with a flat interface. The steady finite-amplitude waves are obtained by solving the full nonlinear inviscid governing equations (Eqs.~\ref{eq:nonlinear_system_eqs}; numerical procedure and non-dimensionalization detailed in \S\ref{sec:Connection to nonlinear waves}) at a fixed non-dimensional wave height (crest-to-trough amplitude) of $h = 0.2$, air-to-water density ratio $r = 0.001$, and Bond number $Bo = 10^4$.
		(a) Regime map of the non-dimensional nonlinear phase speed $\tilde{c}$ normalized by the corresponding irrotational (shear-free) Stokes wave speed $\tilde{c}_S$. Classical irrotational Stokes waves are recovered in the limit $(\tilde{\Delta}, Fr) \rightarrow (0,0)$, while the blank region denotes parameters where our numerical method was unable to find steady solutions.
		(b) Comparison of the non-dimensional air-water interface profile $\tilde{H}_1(\tilde{x})$ (where $\tilde{x}$ is the horizontal coordinate for the PL profile, red solid curves) against the corresponding third-order irrotational Stokes wave (black dashed curves) at the same wave height $h = 0.2$, sampled at the three points marked by magenta circles in (a): (i) $(\tilde{\Delta}, Fr) = (0.8, 0.5)$, (ii) $(\tilde{\Delta}, Fr) = (3.2, 1.6)$, and (iii) $(\tilde{\Delta}, Fr) = (5.2, 2.6)$. Together, (a) and (b) demonstrate that finite-depth background shear substantially increases the nonlinear phase speed and steepens the wave profile (producing sharper crests and flatter troughs) compared to classical irrotational Stokes waves at the same non-dimensional wave height.}
	\label{fig2:Regime_map_stable}
\end{figure}

\begin{figure}[ht!]
	\centering
	\includegraphics[width=0.9\linewidth]{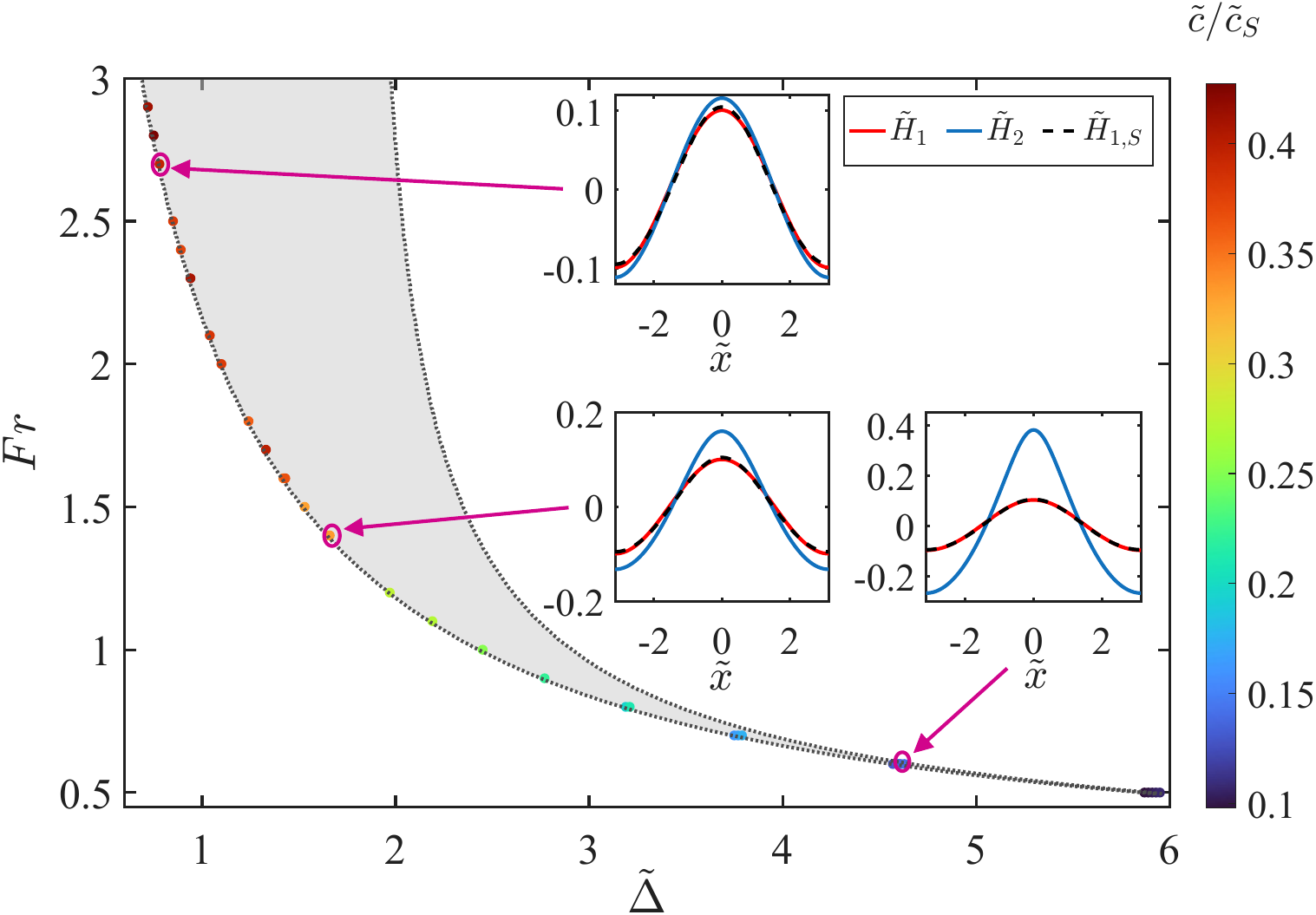}
	\caption{Finite-amplitude gravity-dominated waves. These (each point in the shaded region) represent continuation of infinitesimal amplitude, retrograde Fourier modes (i.e. propagating opposite to the background surface drift, relative to the same), unstable under the flat interface background state (PL profile), computed using the same nonlinear governing equations, numerical procedure (\S\ref{sec:Connection to nonlinear waves}), and physical parameters ($h = 0.2$, $r = 0.001$, $Bo = 10^4$) as in Figure \ref{fig2:Regime_map_stable}.
		The main panel shows the regime map of the non-dimensional nonlinear phase speed $\tilde{c}$ normalized by the corresponding irrotational Stokes wave speed $\tilde{c}_S$, plotted in the parameter space of Froude number $Fr$ and non-dimensional water shear-layer depth $\tilde{\Delta}$. The black dotted curves denote the linear neutral-stability boundaries of the flat interface PL background state. Solutions are reported only in a narrow band near the left neutral-stability boundary due to challenges in the numerical continuation of nonlinear solutions deeper into the unstable regime (shaded region).
		The three insets on the right (connected via magenta arrows to the sampled points marked by magenta circles at $(\tilde{\Delta}, Fr) = (0.78, 2.7)$ [top], $(1.66, 1.4)$ [middle], and $(4.61, 0.6)$ [bottom]) compare the non-dimensional air-water surface wave profile $\tilde{H}_1(\tilde{x})$ (red solid curves) against the corresponding third-order irrotational Stokes wave $\tilde{H}_{1,S}(\tilde{x})$ (black dashed curves) at the same wave height $h = 0.2$, alongside the lower vorticity-interface profile $\tilde{H}_2(\tilde{x})$ (blue solid curves, shifted vertically by $+\tilde{\Delta}$ to share the same axis).
		Across the unstable regime and unlike the prograde waves of fig. \ref{fig2:Regime_map_stable}, finite-amplitude retrograde waves propagate significantly slower than classical Stokes waves ($\tilde{c}/\tilde{c}_S < 1$). While the air-water density interface ($\tilde{H}_1$) closely tracks the irrotational Stokes profile ($\tilde{H}_{1,S}$), the lower vorticity interface ($\tilde{H}_2$) grows increasingly steep and nonlinear with decreasing $Fr$ and increasing $\tilde{\Delta}$.}
	\label{fig3:Regime_map_unstable}
\end{figure}

One of the key results of our study is the demonstration of nonlinear propagating waves which are continuation of infinitesimal amplitude, propagating retrograde modes (see fig. \ref{fig3:Regime_map_unstable} for definition) linearly \textit{unstable}  under the background shear (PL profile) with a flat interface. Before describing these, it is useful to recap the converse regime \citep{milinazzo1990effect,shelton2026nonlinear} i.e. propagating finite-amplitude waves which are continuation of linearly \textit{stable}, infinitesimal amplitude prograde modes for the same background state. Figure \ref{fig2:Regime_map_stable} illustrates such finite-amplitude, propagating surface-gravity waves. These are obtained for the PL profile in the two-dimensional parameter space of Froude number $Fr$ and non-dimensional shear depth $\tilde{\Delta}$. The procedure to obtain such nonlinear propagating waves is also described later on in detail in \S \ref{sec:Connection to nonlinear waves}; figs. \ref{fig2:Regime_map_stable} and \ref{fig3:Regime_map_unstable} present an early overview. The nonlinear propagating waves in fig. \ref{fig2:Regime_map_stable} (each point in the figure describes one such wave, see caption) with the background PL profile, are connected to their  (unsheared) Stokes wave brethren \citep{stokes1847theory}, the latter recovered in the limit $(Fr,\tilde{\Delta})\rightarrow(0,0)$. Their propagation speed (color bar) are presented relative to the unsheared Stokes wave speed, ${\tilde{c}}_S$ (see fig. \ref{fig2:Regime_map_stable} a), and their form compared to the unsheared third-order Stokes wave profile (see fig. \ref{fig2:Regime_map_stable} b). This illustrates that background shear (PL profile) renders the classical unsheared Stokes-wave, steeper and propagating with higher phase speed.  

Figure \ref{fig3:Regime_map_unstable} illustrates finite-amplitude, gravity-dominated propagating waves representing continuation of linearly unstable, infinitesimal amplitude retrograde modes of the sheared background state in the $(Fr,\tilde{\Delta})$ parameter space. In contrast to the waves presented earlier in fig. \ref{fig2:Regime_map_stable}, these finite-amplitude states are connected to the underlying linear instability of the background shear and therefore provide a possible pathway from the infinitesimal amplitude, propagating unstable mode to a nonlinear finite-amplitude propagating state\footnote{note that lacking viscosity in our analysis, the finite amplitude state, when stable, cannot be a true attractor for the exponentially growing solutions from the linear instability.}. These solutions are reported in fig. \ref{fig3:Regime_map_unstable} (points within the shaded region pointed by arrows) only in a limited region of parameter space, close to the left neutral-stability boundary. This restriction arises from the difficulty of continuing the nonlinear solutions across the entire parameter space particularly in the presence of potential bifurcations, a challenge also reported in previous studies of finite-amplitude waves with PL velocity profiles \citep{baumstein1999steady,shelton2026nonlinear}. In \S\ref{subsec:Nonlinear_formulation}, we will circumvent this limitation by restricting ourselves to $Fr=0.8$. As seen from fig. \ref{fig3:Regime_map_unstable}, due to the small width of the unstable regime at this Froude number, it becomes feasible to continue both the marginally unstable as well as the fastest growing linearly unstable mode, into the finite-amplitude regime for this Froude. 

While a weakly nonlinear analysis of the PL profile can provide insight into how the initial exponential growth for the marginally unstable propagating retrograde modes in fig. \ref{fig3:Regime_map_unstable}, can get arrested by nonlinearity, such analyses by definition account only for lowest-order nonlinear effects. To our knowledge, however, the fully nonlinear, unsteady evolution connecting the primary linear instability to a finite-amplitude wave state (of permanent form) in a two-fluid system with a piecewise-linear (PL) shear profile remains unexplored. In the present work, we address this gap by investigating the temporal evolution including initial exponential growth and the subsequent quenching of this for  shear-generated surface gravity waves up to wave breaking. We commence by performing an inviscid linear stability analysis of the background PL profile with a flat air-water interface, in order to quantify the primary unstable modes. 
Building upon these linearised predictions, we then explore the fully nonlinear regime through a dual methodology. First, we carry out high-resolution, numerical simulations (DNS), relying on a two-phase formulation of the Navier-Stokes equations to explicitly resolve the overlying air (in addition to water). This captures the transients of the primary instability, initialised with an infinitesimal amplitude fastest growing linearly unstable mode and eventually reaching a finite-amplitude state. Furthermore, we also formulate an exact, inviscid boundary-value problem with the same background shear (PL profile) to compute the corresponding permanent-form propagating waves admitted by the nonlinear equations. For this, we extend the classical framework of \citet{milinazzo1990effect} by incorporating an active air layer to match the DNS setup, and by permitting fully asymmetric interfacial profiles (details discussed later on in \S\ref{subsec:Nonlinear_formulation}). This degree of freedom is necessary to accommodate the potential phase-shift between the density-vorticity wave (at the air-water interface) and the subsurface vorticity wave (at the lower edge of the shear layer), a feature predicted by linear theory for unstable eigenmodes. 

Our results reveal that for the two-dimensional, coupled air-water system, the fastest growing unstable mode of the background state undergoes exponential growth before ultimately reaching a quenched, finite-amplitude form with further residual growth and distortion prior to breaking. This is characteristic of a supercritical transition, yielding a spatio-temporally complex wave-field dominated by a propagating primary mode and its  phase-locked bound harmonics, while simultaneously also displaying residual free wave components also generated nonlinearly. Crucially, we demonstrate an excellent agreement between the nonlinear states captured through our transient simulations and the exact propagating-wave solutions of permanent form with background shear, computed from our boundary-value formulation. This latter nonlinear theory thus successfully describes broad features of the finite-amplitude interfacial topology and also its phase speed, thereby providing a reliable reference for the pre-breaking nonlinear state for wind-drift instability generated surface gravity waves.

The remainder of this paper is organized as follows. \S \ref{sec:Problem_statement} establishes the physical problem and governing equations. \S \ref{sec:Linear_stability} details the linear stability characteristics of the coupled air-water PL profile. \S \ref{sec:Numerical_simulation} presents the transient numerical simulation setup and wave evolution. \S \ref{sec:Connection to nonlinear waves} derives the finite-amplitude steady-state formulation and compares the exact theoretical solutions against the nonlinear states observed in simulations. Finally, concluding remarks are provided in \S \ref{sec:Conclusions}.

\section{Problem statement}\label{sec:Problem_statement}
\subsection{Physical description}\label{subsec:Physical_description}
\begin{figure}[ht!]
  \centerline{\includegraphics[width = 0.5\textwidth]{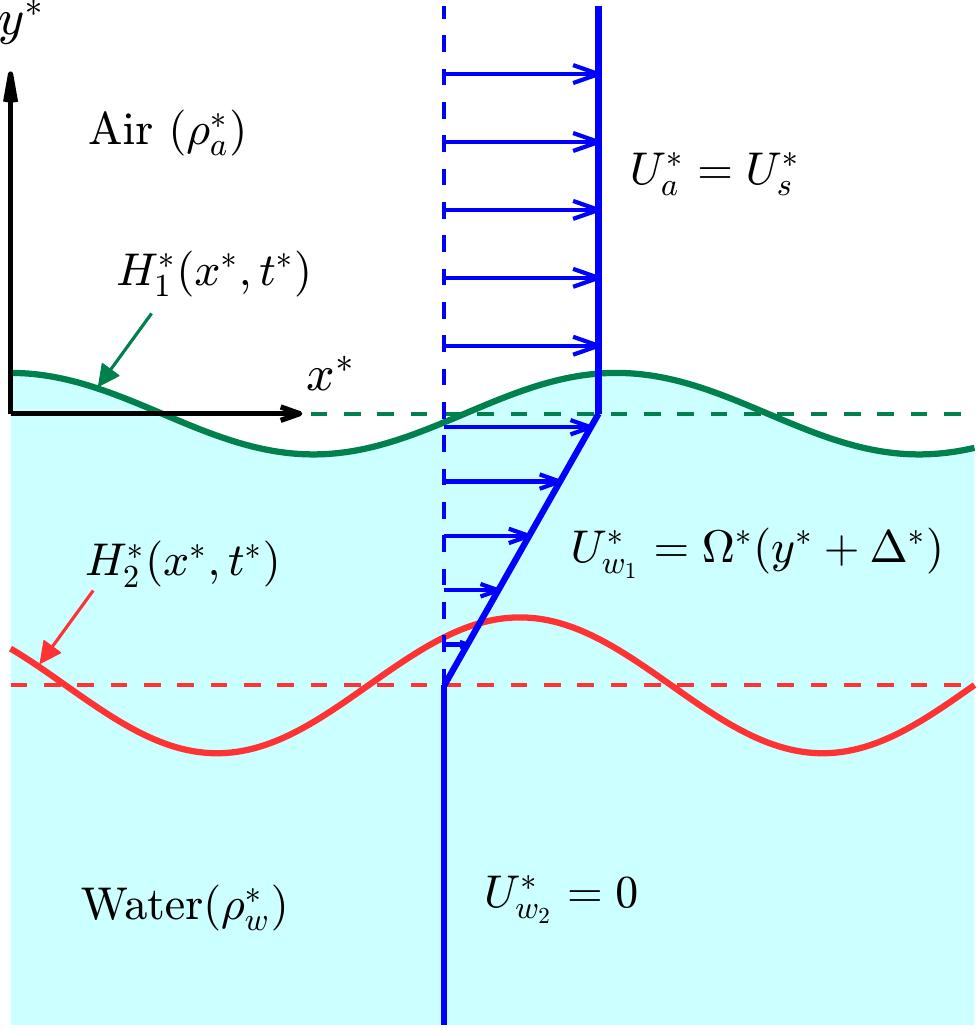}}
  \caption{Schematic of the piecewise-linear (PL) background velocity profile. A uniform air flow ($U_{a}^* = U_s^*$) overlays a finite water shear layer of depth $\Delta^*$ with constant vorticity ($\Omega^* = U_s^*/\Delta^*$), and velocity $U_{w_1}^* = \Omega^*(y^* + \Delta^*)$, which sits above quiescent deep water ($U_{w_2}^* = 0$). The unperturbed boundary at $y^* = 0$ is a density-vorticity interface, whereas the subsurface boundary at $y^* = -\Delta^*$ is a vorticity interface. Their corresponding perturbed forms are denoted by $H_1^*(x^*,t^*)$ and $H_2^*(x^*,t^*)$ respectively. Subscripts $a$, $w_1$, and $w_2$ indicate properties in the air (density $\rho_a^*$), shear layer, and quiescent water (density $\rho_w^*$). }
\label{fig4:Schematic}
\end{figure}
Figure \ref{fig4:Schematic} shows the schematic of a two-dimensional wind-drift driven wave system. Starred variables denote dimensional quantities. The background state comprises a piecewise-linear (PL) profile with air (density $\rho^*_a$) flowing over water (density $\rho^*_w$). The drift profile in water originates from wind exerting a shear stress on the surface, generating a subsurface shear layer of depth $\Delta^*$ characterized by a constant vorticity $\Omega^*$. Because both the overlying uniform air flow and the deep quiescent water ($y^* < -\Delta^*$) are irrotational, this configuration naturally contains two distinct interfacial boundaries. The upper boundary at $y^* = 0$ features a jump in both density and vorticity; hereafter, we refer to this as the density-vorticity interface. The lower boundary at $y^* = -\Delta^*$, which separates the shear layer from the quiescent deep water, is a vorticity interface. The background velocity profile can be expressed as:

\begin{equation}
\bU^*(y^*) = \left\{ \begin{array}{ll}
\displaystyle \bU^*_a(y^*) = U^*_a(y^*)\:\bex =  U^*_s\bex,\quad \quad \quad  \quad \quad \quad \;y^*>0,\\[8pt]
\displaystyle \bU^*_{w_1}(y^*) = U^*_{w_1}(y^*)\:\bex = \Omega^*(y^*+\Delta^*) \:\bex, \quad\;-\Delta^*\leq y^* \leq 0,\\[8pt]
\displaystyle  \bU^*_{w_2}(y^*) = U^*_{w_2}(y^*)\:\bex = \mathbf{0}, \quad \quad \quad\quad \quad \quad \quad\; y^* < -\Delta^*, \end{array}\right.
  \label{eq: base_velocity_PLP}
\end{equation}
which is continuous across the density-vorticity interface, such that $U^*_{w_1}(0) = U^*_s$, which yields $U^*_s = \Omega^*\Delta^*$. Here $\mathbf{\hat{e}}_x$ is an unit vector along the background flow direction.

The perturbed forms of the density-vorticity and subsurface vorticity interfaces are denoted by $H^*_1(x^*,t^*)$ and $H^*_2(x^*,t^*)$ respectively. The properties/flow variables in air, shear layer and quiescent water are denoted by subscripts $a,w_1$ and $w_2$ respectively. 
Both the linear instability and nonlinear propagating wave solutions are obtained for the same background state shown in figure \ref{fig4:Schematic}.

\subsection{Characteristic scales and physical assumptions}

To mathematically formulate the problem, it is instructive to first recap the characteristic physical scales governing the air-water problem. Table \ref{Tab1:physical_scales}(a) summarizes the standard thermophysical properties. Table \ref{Tab1:physical_scales}(b) lists the typical ranges of the characteristic scales in both laboratory wind-wave experiments \citep{wu1975wind, buckley2016structure} and open-ocean measurements \citep{richman1987measurements}. 

\begin{table}
	\begin{center}
		\def~{\hphantom{0}}
		
		\textbf{(a) Thermophysical properties of the air-water system.}\\[1ex]
		\begin{tabular}{lcc}
			Property & Symbol & Value \\[3pt]
			Air density & $\rho^*_a$ & $1.2\text{ kg\,m}^{-3}$ \\
			Water density & $\rho^*_w$ & $1000\text{ kg\,m}^{-3}$ \\
			Kinematic viscosity (air) & $\nu^*_a$ & $1.5 \times 10^{-5}\text{ m}^2\text{s}^{-1}$ \\
			Kinematic viscosity (water) & $\nu^*_w$ & $1.0 \times 10^{-6}\text{ m}^2\text{s}^{-1}$ \\
			Surface tension & $T^*$ & $0.073\text{ N\,m}^{-1}$ 
		\end{tabular}
		\label{Tab1a:properties}
		\vspace{0.8cm}
		
		\textbf{(b) Characteristic scales and their estimates}\\[1ex]
		\begin{tabular}{lcc}
			Parameter & Symbol & Typical Lab (L) / Field (F) Range \\[3pt]
			Free-stream wind speed & $U^*_\infty$ & $2 - 15\text{ m\,s}^{-1}$ (\citealt{wu1975wind}--L; \citealt{richman1987measurements}--F) \\[3pt]
			Interfacial surface velocity & $U^*_s$ & $0.06 - 0.45\text{ m\,s}^{-1}$ (\citealt{phillips1974wave}--L; \citealt{wu1975wind}--L)\\[3pt]
			Subsurface drift layer depth & $\Delta^*$ & $10^{-3} - 10\text{ m}$ (\citealt{wu1975wind}--L; \citealt{richman1987measurements}--F)
		\end{tabular}
		\label{Tab1b:kinematics}
		\caption{Physical scales and governing parameters for the wind-driven air-water interface.}
		\label{Tab1:physical_scales}
	\end{center}
\end{table}

According to the Beaufort scale \citep{stewart2008introduction}, a wind speed of $U^*_{\infty} \sim 2.8\text{ m\,s}^{-1}$ (Beaufort force 2) corresponds to a light breeze, generating ``small wavelets with unbroken, glassy crests''. In contrast, a speed of $U^*_{\infty} \sim 15.8\text{ m\,s}^{-1}$ (Beaufort force 7) corresponds to a near gale, characterized by large waves where white foam from breaking crests begins to be blown into streaks \citep{stewart2008introduction}. A characteristic of such wind-driven interfacial flows is the empirically known relationship between the atmospheric wind speed ($U^*_\infty$) and the induced surface drift velocity ($U^*_s$). As demonstrated by \citet{wu1975wind} and \citet{phillips1974wave}, the surface drift is observed to be approximately $3\%$ of the free-stream wind speed ($U^*_s \approx 0.03 U^*_\infty$) across a wide range of forcing conditions. Consequently, for wind speeds ranging from $2$ to $15\text{ m\,s}^{-1}$, the surface velocity reaches $O(10^{-2})$ to $O(10^{-1})\text{ m\,s}^{-1}$. The momentum from the wind diffuses downwards to form a subsurface drift layer of thickness $\Delta^*$. While this sheared region is typically confined to depths varying from $O(10^{-3})$ to $O(10^{-2})\text{ m}$ in short-fetch laboratory settings, it can extend to $O(1)$ - $O(10)\text{ m}$ or deeper under fully developed oceanic conditions.

Selecting the characteristic length scale as the drift layer thickness ($l_c = \Delta^*$) and the characteristic velocity as the surface drift ($u_c = U^*_s$), the subsurface flow is governed by the water-side Reynolds number,
$$Re = \frac{U^*_s \Delta^*}{\nu^*_w}.$$
Adopting the bounds reported in Table \ref{Tab1:physical_scales}(b), $Re$ ranges from $O(10)$ to $O(10^6)$. Given that $Re \gg 1$ across all physically relevant scenarios, the viscous diffusion timescale is vastly separated from the convective timescale of the wave instability. Consequently, viscosity can be safely neglected in the perturbation equations in a first approximation, allowing the fundamental instability to be evaluated as a inviscid resonance mechanism acting upon a frozen background shear profile (further mathematical justification for the frozen background profile is given in \S \ref{subsec:wave_evolution_linear}).

\subsection{Governing equations and boundary conditions}\label{subsec:Governing_equations}
The flows in air and water are assumed to be incompressible and inviscid and are governed by the continuity and Euler equations as follows ($j = a, w_1, w_2$):
\begin{subequations}
\label{eq: flow_eqs_dimensional}
\begin{align}
    \bnabla^*\bcdot\bu^*_j &= 0, \label{eq: continuity_dimensional}\\
    \frac{\partial \bu^*_j}{\partial t^*} + \bu^*_j\bcdot \bnabla^* \bu^*_j &= -\frac{\bnabla^* p^*_j}{\rho^*_j} + \boldsymbol{g^*}, \label{eq: Euler_eq_dim}
\end{align}
\end{subequations}
Here, $p^*_j$ is the pressure in fluid $j$, and $\boldsymbol{g^*} = -g^*\:\bey$ is the gravitational acceleration, $\bey$ being an upward pointing unit vector along the vertical direction. The base state velocities $\bU^*_j$ and hydrostatic pressure $P^*_j$ satisfy Eq. \eqref{eq: flow_eqs_dimensional}, with $\partial P^*_j/\partial y^* = -\rho^*_j\:g^*$. The flow approaches the base state as $y^*\rightarrow \pm \infty$.

Since the flow is two-dimensional and incompressible, we can therefore define a stream function $\psi^*_j$ for each layer ($j = a, w_1, w_2$) such that it identically satisfies the continuity equation \eqref{eq: continuity_dimensional}:
\begin{equation}
    u^*_j = \partial \psi^*_j/\partial y^*\quad \text{and}\quad v^*_j = -\partial \psi^*_j/\partial x^*. \label{eq: Streamf_dim}
\end{equation}

Kinematic and dynamic boundary conditions are imposed at the perturbed  density-vorticity and vorticity interfaces, $H^*_1(x^*,t^*)$ and $H^*_2(x^*,t^*)$, including surface tension effects:
\begin{subequations} 
\label{eq: BCs_dimensional}
\begin{align}
\frac{\partial F^*_s}{\partial t^*} + \bu^*_s&\bcdot\bnabla^* F^*_s = 0, \quad &\text{at}: y^* &= H^*_s(x^*, t^*), \quad (s=1,2), \label{eq: KBC_dimensional} \\
p^*_{w_1} - p^*_a &= T^*\:(\bnabla^* \bcdot \bn_1), \quad &\text{at}: y^* &= H^*_1(x^*, t^*), \label{eq: normal_stress_BC_dimensional_H1}\\
p^*_{w_1} - p^*_{w_2} &= 0, \quad &\text{at}: y^* &= H^*_2(x^*, t^*), 
\label{eq: normal_stress_BC_dimensional_H2}
\end{align}
\end{subequations}
where $F^*_s(\bx^*, t^*) = y^* - H^*_s(x^*, t^*) = 0$ defines the material surfaces, $T^*$ is the surface tension, and $\bn_1 = \bnabla^* F^*_1/|\bnabla^* F^*_1|$ is the unit normal to the density-vorticity interface. The kinematic conditions \eqref{eq: KBC_dimensional} enforce that fluid particles cannot cross the interfaces.
Equations \eqref{eq: flow_eqs_dimensional}, \eqref{eq: Streamf_dim} subjected to \eqref{eq: BCs_dimensional} form the system of equations for our problem. These equations are linearised about the PL background state and details of the linear stability analysis are presented subsequently.

\section{Linear stability analysis}
\label{sec:Linear_stability}
\subsection{Important non-dimensional numbers}
\label{subsec:LSA_char_scales}
Before formulating the temporal linear stability problem, it is useful to non-dimensionalize the governing equations \eqref{eq: flow_eqs_dimensional}-\eqref{eq: BCs_dimensional} using the surface air velocity and the shear layer depth as our primary reference scales. Table \ref{tab:Tab2_scales_LSA} lists the characteristic scales chosen for this analysis. 
\begin{table}
  \begin{center}
\def~{\hphantom{0}}
  \begin{tabular}{lcl}
      Variable  & Characteristic scale   &   Description \\[3pt]
       Length   & $\Delta^*$ & Shear layer depth \\
       Velocity   & $U^*_s$ & Uniform air velocity \\
       Time  & $\Delta^*/U^*_s$ & Advection time scale \\
       Stream function & $\Delta^* U^*_s$ & Kinematic flow scale \\
       Pressure & $\rho^*_w U^{*2}_s$ & Convective pressure scale\\
       Density & $\rho^*_w$ & Quiescent water density
  \end{tabular}
  \caption{Characteristic scales for the linear stability analysis.}
  \label{tab:Tab2_scales_LSA}
  \end{center}
\end{table}
Applying these yields three primary non-dimensional parameters that govern the linear instability: the air to water density ratio $r$, the Froude number $Fr$, characterized by ratio of inertia to gravity and the Bond number $Bo$, characterized by the ratio of gravity to surface tension. Precise definitions of these along with their respective ranges (ocean) are given in table \ref{tab:Tab3_nondim_numbers_LSA}. 
\begin{table}
  \begin{center}
\def~{\hphantom{0}}
  \begin{tabular}{lcl}
      Dimensionless number  & Definition   &   Value range \\[3pt]
       Density ratio ($r$)   & $\rho^*_a/\rho^*_w$ & $0.001$ \\
       Froude number ($Fr$)   & $U^*_s/\sqrt{g^* \Delta^*}$ & $O(10^{-3}) - O(1)$ \\
       Bond number ($Bo$)  & $\rho^*_w g^* \Delta^{*2}/T^*$ & $O(10^{-1}) - O(10^7)$
  \end{tabular}
  \caption{Key dimensionless numbers with their range of values, employing values in table \ref{Tab1:physical_scales}.}
  \label{tab:Tab3_nondim_numbers_LSA}
  \end{center}
\end{table}
We drop the asterisks hereafter working with non-dimensional variables for the remainder of this section.

\subsection{Temporal linear stability formulation}
\label{sec:LSA_formulation}

We introduce infinitesimal, two-dimensional perturbations to the base state using a standard normal mode expansion, such that the total stream function is given by $\psi_j(x, y, t) = \Psi_j(y) + \hat{\psi}_j(y)\exp\big\{{\mathrm{i} k(x - c t)}\big\}$. Here, $\mathrm{i} = \sqrt{-1}$, while $k$ and $c$ are the non-dimensional wavenumber and complex phase speed, respectively. Substituting this decomposition into the governing equations established in \S \ref{subsec:Governing_equations} and linearizing yields the classic Rayleigh equation for the perturbation amplitude in each layer ($j = a,w_1,w_2$):
$(U_j(y)-c)\left[\mathcal{D}^2 - k^{2}\right]\hat{\psi}_j(y)  - \left[\mathcal{D}^2 U_j(y)\right]\hat{\psi}_j(y) = 0,$
where $\mathcal{D} \equiv \mathrm{d}/\mathrm{d}y$. Because the base velocity profile is piecewise-linear, $\mathcal{D}^2 U_j = 0$ everywhere except at the interfaces, where it has Dirac delta function behaviour. Consequently, within each fluid layer, the Rayleigh equation simplifies to:
\begin{equation}
    \left[\mathcal{D}^2 - k^{2}\right]\hat{\psi}_j(y) = 0. \label{eq: Rayleigh_LSA_nondim_simplified}
\end{equation}
From \eqref{eq: Rayleigh_LSA_nondim_simplified} and the requirement that perturbations vanish in the far-field ($y \to \pm\infty$), the general solutions for the perturbation stream functions in the three layers are:
\begin{subequations} \label{eq: psi_sol_LSA}
    \begin{align}
        \hat{\psi}_a(y) &= A\exp(-ky), \quad &\text{for}\quad &y \geq 0, \\[6pt]
        \hat{\psi}_{w_1}(y) &= B\:\exp(ky) + D\:\exp(-ky), \quad &\text{for}\quad -1 \leq \;&y < 0, \\[6pt]
        \hat{\psi}_{w_2}(y) &= E\:\exp(ky), \quad &\text{for}\quad &y < -1.
    \end{align}
\end{subequations}
where $A,B,D,E$ are (complex) constants of integration. The perturbed density-vorticity and vorticity interfaces can be similarly represented as $y = H_1(x,t) = \eta_d\:\exp\big\{{\mathrm{i} k(x - c t)}\big\}$ and $y=H_2(x,t) = -1 + \eta_v\:\exp\big\{{\mathrm{i} k(x - c t)}\big\}$. The perturbed pressure is represented as $p_j(x,y,t) = P_j(y) + \hat{p}_j(y)\:\exp\big\{{\mathrm{i} k(x - c t)}\big\}$, where the perturbation amplitude is related to the streamfunction by:
\begin{equation}
    \hat{p}_j(y) = \rho_j\left[-(U_j(y)-c)\left(\mathcal{D}\hat{\psi}_j(y)\right) + \left(\mathcal{D}U_j(y)\right)\hat{\psi}_j(y)\right]. \label{eq:Pressure_LSA_nondim}
\end{equation}
Substituting these expansions into the nonlinear boundary conditions \eqref{eq: BCs_dimensional} and linearizing about the undisturbed interface locations ($y=0$ and $y=-1$) yields the following linearised system:
\begin{subequations}
    \begin{align}
        &(c - 1)\:\eta_d  = \hat{\psi}_a(0) = \hat{\psi}_{w_1}(0), \label{eq: KBC_nondim_H1} \\[6pt]
        &c\:\eta_v  =  \hat{\psi}_{w_1}(-1) = \hat{\psi}_{w_2}(-1), \label{eq: KBC_nondim_H2} \\
        &-(1-c)\:\mathcal{D}\hat{\psi}_{w_1}(0) + \hat{\psi}_{w_1}(0) + r\:(1-c)\:\mathcal{D}\hat{\psi}_a(0) = \eta_d\left[\frac{1-r}{Fr^2} + \frac{k^{2}}{Fr^2Bo}\right], \label{eq: normal_stress_BC_nondim_H1} \\
        &c\:\mathcal{D}\hat{\psi}_{w_1}(-1) + \hat{\psi}_{w_1}(-1) = c\:\mathcal{D}\hat{\psi}_{w_2}(-1). \label{eq: normal_stress_BC_nondim_H2}
    \end{align}
    \label{eq: BCs_LSA_nondim}
\end{subequations}

Substituting the general solutions \eqref{eq: psi_sol_LSA} into the boundary conditions \eqref{eq: BCs_LSA_nondim} and requiring a non-trivial solution yields a cubic dispersion relation for the complex phase speed $c(k)$:
\begin{equation} \label{eq: dispersion_relation_LSA_nondimensional}
    a_3 c^3 + a_2 c^2 + a_1 c + a_0 = 0,
\end{equation}
where the coefficients are given by:
\begin{subequations}
    \begin{align*}
        a_3 &= 2k^2(1 + r), \\[6pt]
        a_2 &= -k\Big[(1 - r)Q + (-1 + r + 4k(1 + r))\Big], \\[6pt]
        a_1 &= -\frac{2k(1-r)}{Fr^2} - \frac{2k^3}{Bo\, Fr^2} + \Big[-1 + Q\big(1 - 2k(-1 + r)\big) + 2k\big(r + k(1 + r)\big)\Big], \\[6pt]
        a_0 &= 1 - Q - k\Big(Q(1 - r) + 1 + r\Big) - \frac{(1 - Q)(r - 1)}{Fr^2} + \frac{k^2(1-Q)}{Bo\, Fr^2},
    \end{align*}
\end{subequations}
with $Q \equiv  \mathrm{e}^{-2k}$. For a given (real) wavenumber $k$, $Fr$ and $Bo$ complex roots of cubic polynomial \eqref{eq: dispersion_relation_LSA_nondimensional} are computed numerically to identify the unstable modes. The functional forms of $\eta_v, A,B,D,E$ in terms of $\eta_d$ (free variable), $k,c$ are reported in Appendix \ref{appA:complex_coefficients_LSA}. Validation of the linear stability formulation against the results of \citet{morland1991waves} is provided in Appendix \ref{appA:validation_LSA}. 

\subsection{Linear stability characteristics}
\begin{figure}[ht!]
  \centerline{\includegraphics[width = \textwidth]{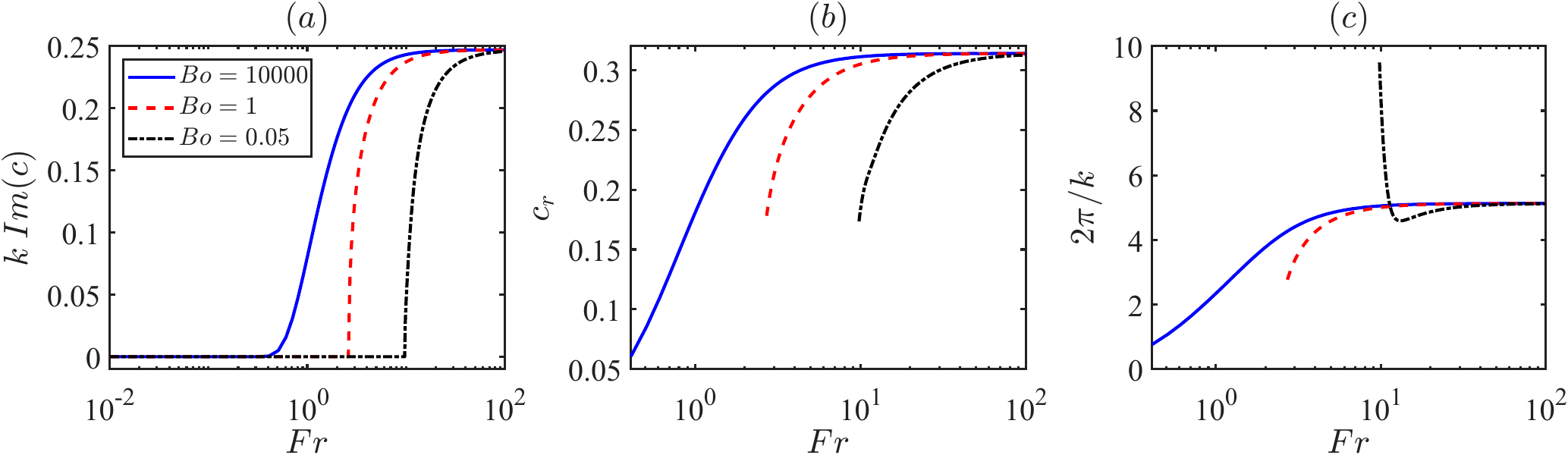}}
  \caption{Linear stability characteristics of the most unstable (fastest growing) mode as a function of the Froude number ($Fr$). The panels display: (a) growth rate, $k\,\mathrm{Im}(c)$ with $\mathrm{Im}(c)$ denoting the imaginary part of $c$; (b) real part of the phase speed, $c_r \equiv \mathrm{Re}(c)$; and (c) wavelength, $2\pi/k$. Results are obtained by solving the dispersion relation \eqref{eq: dispersion_relation_LSA_nondimensional} for an air-water density ratio of $r = 0.001$. The curves correspond to different Bond numbers: $Bo = 10^4$ (solid line), $Bo = 1$ (dashed line), and $Bo = 0.05$ (dash-dotted line).}
\label{fig5:LSA_dispersion}
\end{figure}

Figure \ref{fig5:LSA_dispersion} depicts the (temporal) linear stability characteristics of the background state (figure \ref{fig4:Schematic}), obtained by solving the dispersion relation \eqref{eq: dispersion_relation_LSA_nondimensional} in MATLAB \citep{MATLAB:R2025a} for an air-water density ratio of $r = 0.001$ and chosen values of wavenumber $k$, Bond number $Bo$ and Froude number $Fr$. We track the growth rate (in panel (a)), phase speed (in panel (b)), and wavelength (in panel (c)) of the most unstable mode as a function of the Froude number ($Fr$), for different choices of Bond numbers $Bo = 10^4,1$ and $0.05$.

As such, we observe a clear threshold behavior, where, the flow remains stable until a critical Froude number ($Fr_{cr}$) is exceeded, after which the growth rate $k\,\mathrm{Im}(c)$ rises steeply. Since $Fr \propto U^*_s/\sqrt{\Delta^*}$, this instability threshold can be crossed either through an increase in the free-stream wind speed, or a thinning of the subsurface shear layer \citep{morland1991waves}. In the limit of negligible surface tension ($Bo = 10^4$), this onset occurs at $Fr_{cr} \approx 0.3$. The influence of surface tension is highly stabilizing; as the Bond number decreases (indicating stronger surface tension), the onset of instability is delayed to progressively higher Froude numbers. However, at sufficiently high Froude numbers ($Fr \gg 1$), achieved through intense wind forcing or very thin shear layers, the growth rates for all Bond numbers converge to an asymptotic limit, demonstrating that capillary effects become negligible. 

This transition is mirrored in panels (b) and (c) of fig. \ref{fig5:LSA_dispersion}. For the gravity-dominated regime as well as moderate strength of capillarity (relative to gravity), both the phase speed $c_r$ and the wavelength $2\pi/k$ of the fastest growing mode rapidly increase near the stability boundary before plateauing at large $Fr$, reflecting a transition toward longer and faster waves. Conversely, in the strongly capillary-dominated regime ($Bo = 0.05$), see dashed-dot curve in panel (c) of fig. \ref{fig5:LSA_dispersion}, the instability commences at relatively longer wavelengths at $Fr = Fr_c$, sharply decreasing with $Fr > Fr_c$ and ultimately converging to the high-$Fr$ plateau. 

\subsection{Instability mechanism: wave interaction perspective}

To develop a physical picture of the mechanism of instability discussed in the previous section, it is instructive to first consider \textit{in isolation}, the wave modes supported by each of the three interfaces (vorticity and density interfaces at $y=0$ and the vorticity interface at $y=-1$) which appear in our background state \citep{carpenter2011instability}. In the absence of background shear, the unperturbed density interface at $y = 0$ supports stable prograde and retrograde capillary-gravity waves, with phase speeds denoted by $c_g^+$ and $c_g^-$, respectively. Similarly, in the absence of density stratification at $y=0$ and without the kink in the background profile at $y=-1$, there is only a vorticity interface at $y=0$, supporting a stable retrograde vorticity wave with phase speed $c_v(0)$. Analogously, the isolated subsurface vorticity interface at $y = -1$ (sans any density stratification at $y=0$), supports a stable prograde vorticity wave with phase speed $c_v(-1)$. The analytical expressions for the phase speeds of these isolated waves are given by \citep{carpenter2011instability}:
\begin{subequations} \label{eq: cg_cv_formulae}
    \begin{align}
        c^{\pm}_g &= 1 \pm \frac{1}{Fr}\sqrt{\frac{1-r}{1+r}\left(\frac{1}{k}\right) + \frac{1}{1+r}\left(\frac{k}{Bo}\right)}, \\[6pt]
        c_v(0) &= 1 - \frac{1}{2k}, \quad  c_v(-1) = \frac{1}{2k}.
    \end{align}
\end{subequations}

\begin{figure}[ht!]
  \centerline{\includegraphics[width = \textwidth]{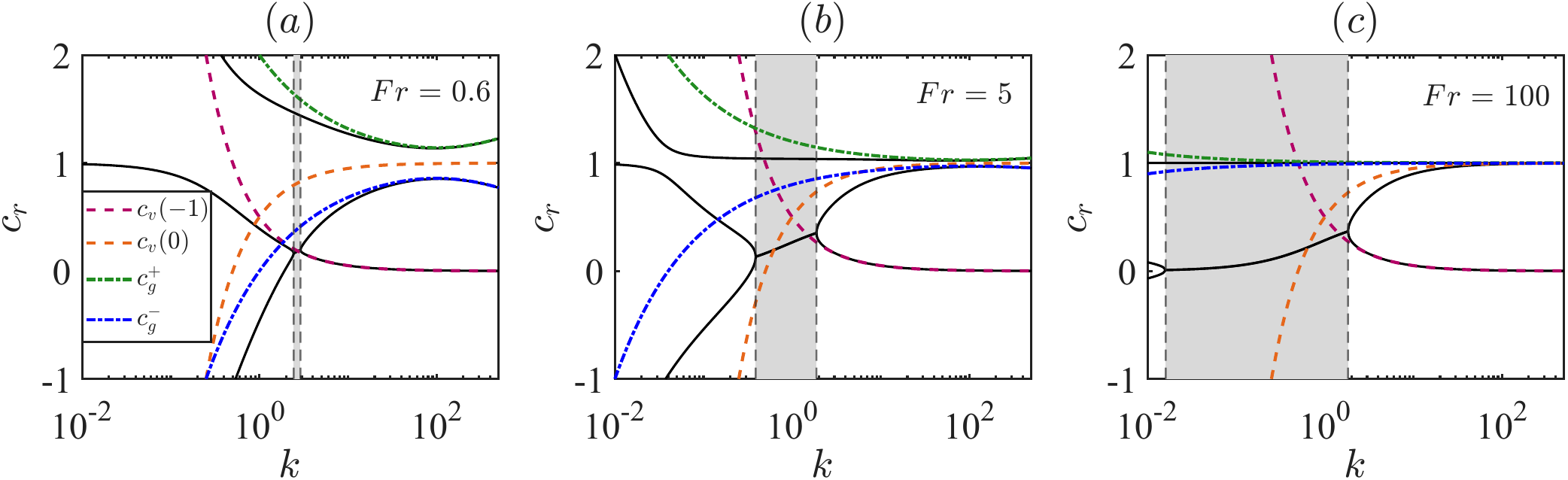}}
  \caption{Phase speeds $c_r$ of the interacting and isolated wave modes as a function of wavenumber $k$ for (a) low ($Fr = 0.6$), (b) moderate ($Fr = 5$), and (c) high ($Fr = 100$) Froude numbers. Solid black curves represent the exact roots of the full dispersion relation \eqref{eq: dispersion_relation_LSA_nondimensional}. Overlaid are the non-interacting phase speeds defined in \eqref{eq: cg_cv_formulae}: the prograde ($c_v(-1)$, dashed magenta) and retrograde ($c_v(0)$, dashed orange) vorticity waves, alongside the prograde ($c_g^+$, dash-dotted green) and retrograde ($c_g^-$, dash-dotted blue) capillary-gravity waves. The gray shaded regions indicate the unstable wavenumber bands where modes phase-lock and $c$ becomes complex. Fixed parameters are $Bo = 10^4$ and $r = 0.001$.}
\label{fig6:LSA_wave_interaction}
\end{figure}

Figure \ref{fig6:LSA_wave_interaction} illustrates the variation of the phase speeds $c_r$ with wavenumber $k$, overlaying the exact interacting modes (from the cubic dispersion relation \eqref{eq: dispersion_relation_LSA_nondimensional}) along with the isolated modes defined in \eqref{eq: cg_cv_formulae}. The onset of instability occurs when the phase-speed ($c_r$) curves for two counter-propagating modes intersect (see intersecting black curves in panel (a)). Such modes can phase-lock and travel at a common intermediate speed, allowing them to resonantly amplify within the shaded wavenumber bands in the figure - this is the linear instability of our interest. The physical origin of these intersecting branches on the $c_r-k$ plane can now be identified by tracing their isolated counterparts outside the unstable regions, as seen in fig. \ref{fig6:LSA_wave_interaction}.

In figure \ref{fig6:LSA_wave_interaction}(a), at low wind forcing ($Fr = 0.6$), the instability is governed by a Holmboe-type resonance. The unstable wavenumber band emerges where the retrograde capillary-gravity wave ($c_g^-$) intersects and resonates with the prograde subsurface vorticity wave ($c_v(-1)$). Conversely, under intense wind forcing ($Fr = 100$, panel c), a qualitatively different behaviour is observed. The capillary-gravity modes become passive, and the instability is instead dominated by a Kelvin-Helmholtz-type resonance between the upper retrograde vorticity wave ($c_v(0)$) and the lower prograde vorticity wave ($c_v(-1)$), resulting in a broad band of instability that extends to long wavelengths. Bridging these extremes is the moderate wind forcing regime ($Fr = 5$, panel b), which acts as a transitional state where a hybrid of both Holmboe and Kelvin-Helmholtz resonance mechanisms co-exist. It is important to note that the transition seen in the nature of the instability in fig. \ref{fig6:LSA_wave_interaction} with increasing $Fr$, is a smooth one.

\section{Numerical simulations and nonlinear evolution of the Holmboe instability}
\label{sec:Numerical_simulation}
\subsection{Numerical setup}
To investigate the nonlinear evolution of the shear-driven Holmboe instability studied by linear theory so far, we perform two-dimensional numerical simulations initialized with the most unstable eigenmode identified in \S \ref{sec:Linear_stability}, at low $Fr$ and high Bond number. These simulations are conducted using the open-source solver Basilisk \citep{popinet2025basilisk}. The two-phase flow is modeled using a one-fluid Volume-of-Fluid (VoF) formulation, which implicitly captures and advects the sharp air-water interface. We define a volume fraction field, $f(x^*,y^*,t^*)$ (where the superscript $^*$ denotes dimensional quantities), which equals $1$ in the water phase and $0$ in the air phase. Transitional values of $f \in (0,1)$ identify computational cells containing the interface. The precise location of the air-water interface ($H_1^*$) is extracted from the $f = 0.5$ isocontour here. The evolution of the volume fraction field is governed by the advection equation:
\begin{equation}
    \frac{\partial f}{\partial t^*} + \bu^*\bcdot\bnabla^* f = 0. \label{eq: vof_advection}
\end{equation}
The fluid velocity $\bu^*(x^*,y^*,t^*)$ and pressure $p^*(x^*,y^*,t^*)$ fields are governed by the continuity and Navier-Stokes equations respectively:
\begin{subequations}
\label{eq: flow_eqs_basilisk}
\begin{align}
    \bnabla^*\bcdot\bu^* &= 0, \label{eq: continuity_basilisk}\\[6pt]
    \frac{\partial \bu^*}{\partial t^*} + \bnabla^* \bcdot (\bu^*\otimes\bu^*) &= \frac{1}{\rho^*} \left[-\bnabla^* p^* + \bnabla^*\bcdot(2\mu^*\bD^*)\right] + \boldsymbol{g^*} + \frac{T^*}{\rho^*}\kappa^*\delta_s\bn, \label{eq: NS_eq_basilisk}
\end{align}
\end{subequations}
where $\boldsymbol{g^*}$ is the gravitational acceleration vector, $\mu^*$ is the dynamic viscosity, and $\bD^* = [\bnabla^*\bu^* + (\bnabla^*\bu^*)^T]/2$ is the rate-of-strain tensor. The final term in \eqref{eq: NS_eq_basilisk} represents the surface force, where $T^*$ is the surface tension coefficient, $\kappa^*$ is the interfacial curvature, $\bn$ is the unit normal vector, and $\delta_s$ is a Dirac delta function at the interface. The local density $\rho^*$ and viscosity $\mu^*$ are computed as arithmetic averages of values of the two phases, weighted by the local volume fraction $f$. Recall from \eqref{eq: base_velocity_PLP} that the unperturbed piecewise-linear base flow possesses a constant vorticity $-\Omega^*$ within the subsurface shear layer, bounded by irrotational fluid. Therefore, we track the perturbed subsurface vorticity interface ($H_2^*$) by extracting the $-\Omega^*/2$ vorticity isocontour (the average of zero vorticity below the shear layer and $-\Omega^*$ in the shear layer). It has been checked that this definition for the lower vorticity interface agrees well when overlaid on the contours of the total vorticity field in simulations, except very close to breaking conditions. Similarly, the contour of vorticity for the upper interface agrees well with that of the density interface even in the nonlinear regime. In the following, thus, we will report the upper interface $H_1^*$ as obtained from the density interface in our numerical simulations.

Throughout the numerical investigation, we maintain the air-water density ratio at $r = 0.001$ and operate in the gravity-dominated regime $(Bo = 10^4)$ (see table \ref{tab:Tab3_nondim_numbers_LSA}). While the fundamental wave dynamics of the instability are inherently inviscid, purely inviscid numerical formulations (Euler equations) for interfacial flows can be susceptible to numerical noise upon small scale grid refinement e.g. see discussion in Appendix C in \cite{kayal2025focussing}. To eliminate these while also excluding damping of the primary physical mechanisms of interest, a small viscosity is introduced thus solving the Navier-Stokes equations. These kinematic viscosities ($\nu_{w}^* = 7.5\times10^{-5} \text{ cm}^2 \text{ s}^{-1}$ for water and $\nu_{a}^* = 1.125 \times 10^{-3} \text{ cm}^2 \text{ s}^{-1}$ for air) are scaled so as to match the kinematic viscosity ratio of $\nu_a^*/\nu_w^* = 15$, identical to air and water at standard conditions in table \ref{Tab1:physical_scales}a. Furthermore, this scaling yields a macroscopic Reynolds number of $Re \approx 10^5$, ensuring that our numerical computations maintain dynamic similitude with the physical high-Reynolds-number inviscid regime of current interest, while also eliminating artifacts associated with fine-scale grid refinement of purely inviscid simulations.

\begin{figure}[ht!]
  \centerline{\includegraphics[width = 0.65\textwidth]{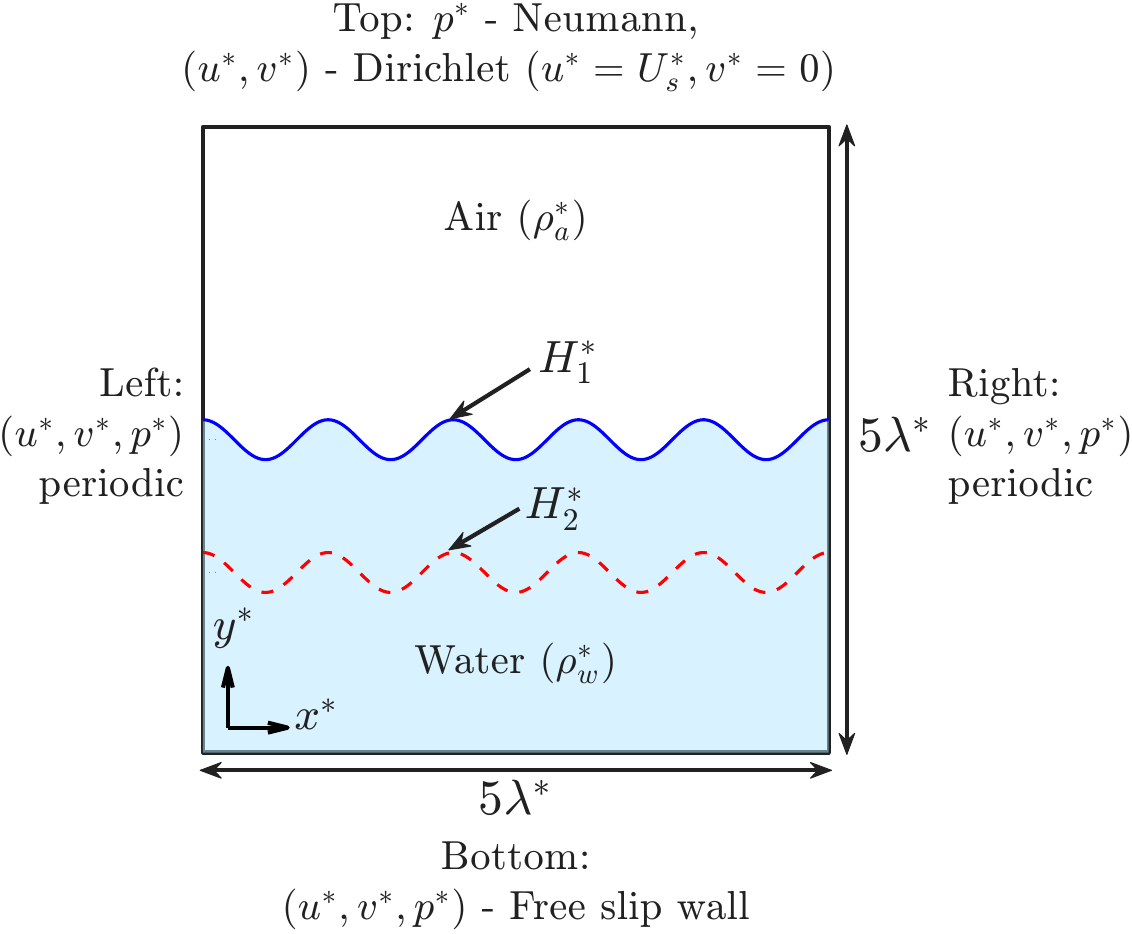}}
  \caption{Schematic of the two-dimensional computational domain and boundary conditions implemented.}
 \label{fig7:numerical_setup}
\end{figure}
Figure \ref{fig7:numerical_setup} illustrates the two-dimensional computational domain and the corresponding boundary conditions. The flow is initialized using the most unstable eigenmode extracted from the linear stability analysis (\S \ref{sec:Linear_stability}). Specifically, the system begins with the air ($\rho_a^*$) and water ($\rho_w^*$) phases separated by the perturbed density-vorticity interface $H_1^*(x^*,0) = \mathrm{Re} \{\eta_d^* \mathrm{e}^{\mathrm{i} k^*x^*} \}$\footnote{Re$\left(\cdot\right)$ indicates real part of the argument} and the subsurface vorticity interface $H_2^*(x^*,0) = -\Delta^* + \mathrm{Re} \{\eta_v^* \mathrm{e}^{\mathrm{i} k^*x^*} \}$. The initial velocity field across all three layers is reconstructed from the eigenfunctions \eqref{eq: psi_sol_LSA} mapped to the dimensional physical domain using \eqref{eq: Streamf_dim}. To allow for the nonlinear wave amplification, the square computational domain in fig. \ref{fig7:numerical_setup} spans an area of $5\lambda^* \times 5\lambda^*$, where $\lambda^*$ is the wavelength of the primary unstable mode. Analytical justification for the chosen domain size, derived from the far-field decay of the nonlinear state, is provided in Appendix \ref{appB:domain_size}. All simulations are initialized with a small wave steepness ($\eta_d^*k^* = 0.02$) to satisfy the assumptions of linear theory.

To resolve the interfaces, we employ the quadtree adaptive mesh refinement (AMR) scheme in Basilisk \citep{popinet2025basilisk}. The spatial resolution is continuously updated based on local variations in the volume fraction ($f$), the interfacial curvature ($\kappa^*$), and the velocity vector components ($u^*, v^*$). The absolute error thresholds are set to $10^{-3}$ for $f$, $10^{-4}$ for $\kappa^*$, and $5 \times 10^{-3}$ for the velocity field. By adapting on $f$ and $\kappa^*$, the solver continuously allocates maximum spatial resolution at the density-vorticity interface ($H_1^*$). Simultaneously, adapting on the velocity components ensures that the steep velocity gradients within the subsurface shear layer ($H_2^*$) are also captured, while grids in the homogeneous far-field irrotational regions remain relatively coarse. The grid resolution is bounded between a minimum refinement level of $7$ ($2^7$ is the upper limit of grid points along each direction) and a maximum level of $12$. The smallest computational cell size in the domain is thus $5\lambda^*/2^{12}$. 

\subsection{Linear to nonlinear regimes and breaking}\label{subsec:energy}
\begin{figure}
	\centerline{\includegraphics[scale=0.45]{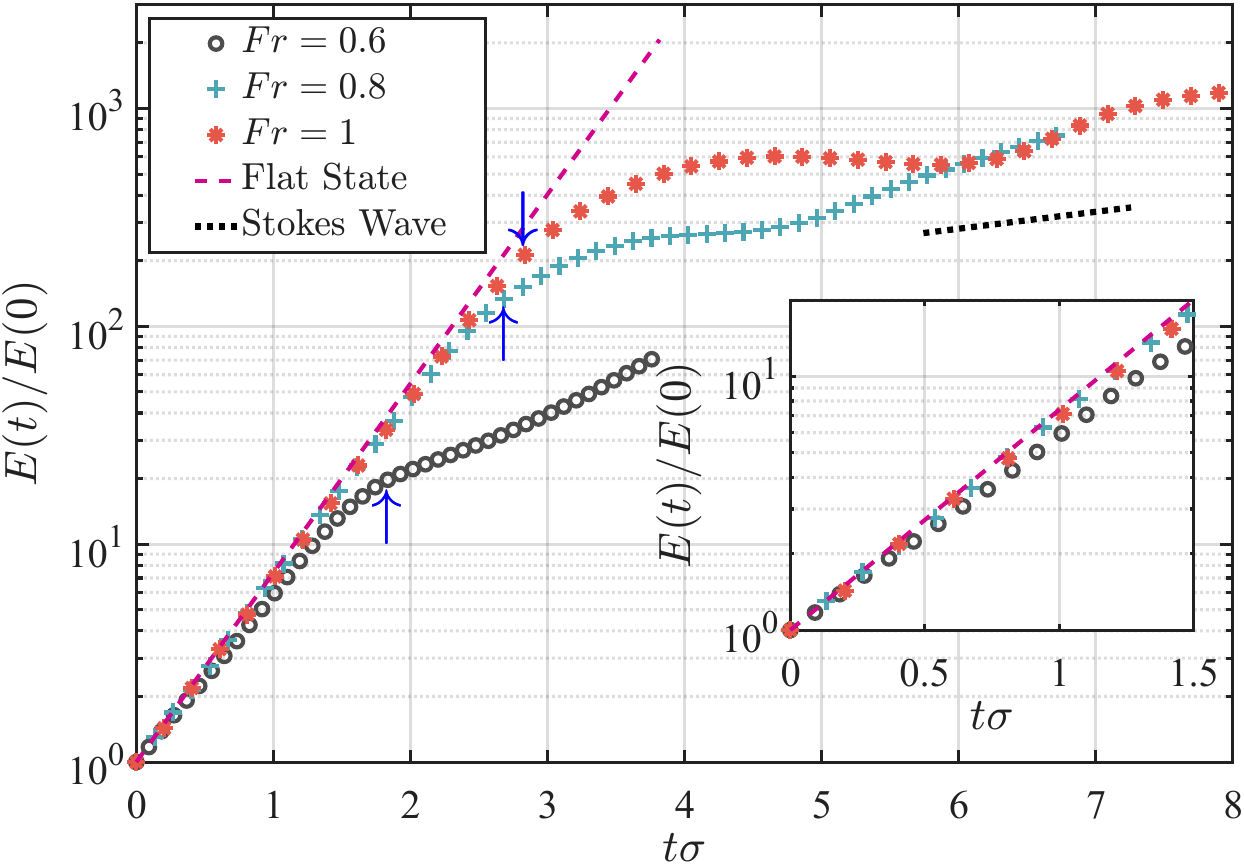}}
	\caption{Temporal evolution of the normalized potential energy at the density interface, $E(t)/E(0)$, versus scaled time $t\sigma$ (where $\sigma = k c_i$ is the linear growth rate) on a semi-log scale for $Fr = 0.6$ (gray circles), $0.8$ (sky blue pluses), and $1$ (orange asterisks), at fixed $Bo = 10^4$ and $r = 0.001$. For all $Fr$, the initial exponential growth matches the linear stability prediction for the background state PL profile with a flat interface (dashed magenta line, labeled ``Flat State''; see inset for a close-up of $0 \le t\sigma \le 1.5$). Blue arrows mark the onset of the submerged vorticity interface breaking for each data-set. The data-sets for $Fr = 0.8$ and $1$ are shown up to the onset of density interface breaking for either cases, whereas no such density interface breaking was observed for $Fr = 0.6$ up to the simulated time reported ($t\sigma \approx 3.8$). The secondary growth stage of the finite-amplitude waves can be compared against the exponential growth rate of the fastest growing mode (subharmonic perturbation) for the linearly unstable, unsheared Stokes wave (black dotted line, labeled ``Stokes Wave'') \citep{longuet1978subharmonic}.}
	\label{fig8:Wave_energy}
\end{figure}
In nonlinear simulations, the evolution of the density-vorticity as well as the submerged vorticity interface, progresses through multiple transitions. To quantify these, we employ as a metric, the growth-rate of the gravitational potential energy of the density interface ($H^*_1(x^*,t^*)$), calculated in the same way as \cite{wu2022revisiting}:
\begin{equation}\label{eq:Energy}
	E(t) \equiv \rho^*_w g^* L^* \int_{-L^*/2}^{L^*/2} (H_1^*(x^*,t^*))^2 \, dx^*.
\end{equation}
where $L^* = 5\lambda^*$. Figure~\ref{fig8:Wave_energy} reports $E(t)$ against time as obtained from simulations, the time-axis for each data-set scaled by its linear growth rate ($t\sigma$, where $\sigma \equiv kc_i$, $c_i$ being the imaginary part of the phase-speed obtained from the dispersion relation \ref{eq: dispersion_relation_LSA_nondimensional}) for $Fr = 0.6, 0.8$, and $1$, for the fastest growing mode in each case. The wave evolution is seen to follow three primary stages: (a) a linear regime exhibiting exponential growth (\S\ref{sec:Linear_stability}), all three date-sets collapse onto a single curve (dashed magenta line) in fig. ~\ref{fig8:Wave_energy} within this regime (b) onset of nonlinear regime where the energy growth falls below its linearised prediction (inset indicates this time is $Fr$ dependent and occurs earliest for $Fr=0.6$) and culminating in the breaking of the submerged vorticity interface (see blue arrows) and (c) a third regime where a brief plateau followed by secondary growth, is discernible for $Fr=0.8$ and $Fr=1$ culminating in the breaking of the density interface for both; beyond this time data is not reported, note that density interface breaking occurs only for $Fr=0.8,1$ but not for $Fr=0.6$. The transition between deviation from linearised exponential growth (stage (b) above) and the onset of the breaking of the vorticity interface (blue arrows in fig. ~\ref{fig8:Wave_energy}) define our target ``\textit{nonlinear window}'' for our finite-amplitude analysis, reported later on in \S\ref{subsec:boundwaves}. This time window commences at the instant when the potential-energy deviates by $15\%$ from its linearised prediction, and ends when the subsurface vorticity wave begins to break (blue arrows in fig. ~\ref{fig8:Wave_energy}). That the submerged vorticity wave breaking significantly precedes that of the density wave may be understood from the fact that the former is steeper than the latter in linear theory. The submerged vorticity wave is thus already highly nonlinear at the start of the nonlinear window. The growth rate of the fastest growing mode (long wave subharmonic mode) for the linearly unstable, unsheared Stokes wave, is also indicated as the back-dotted line in fig. \ref{fig8:Wave_energy}; this was calculated using the theory of \cite{longuet1978subharmonic}. This serves as a plausible reference to justify the observed growth in wave energy following the brief plateau for the cases $Fr=0.8,1.0$ in fig. \ref{fig8:Wave_energy}.

\subsection{Spatio-temporal evolution of $H_1$ and $H_2$} \label{subsec:wave_evolution_linear}
\begin{figure}
  \centerline{\includegraphics[width = \textwidth]{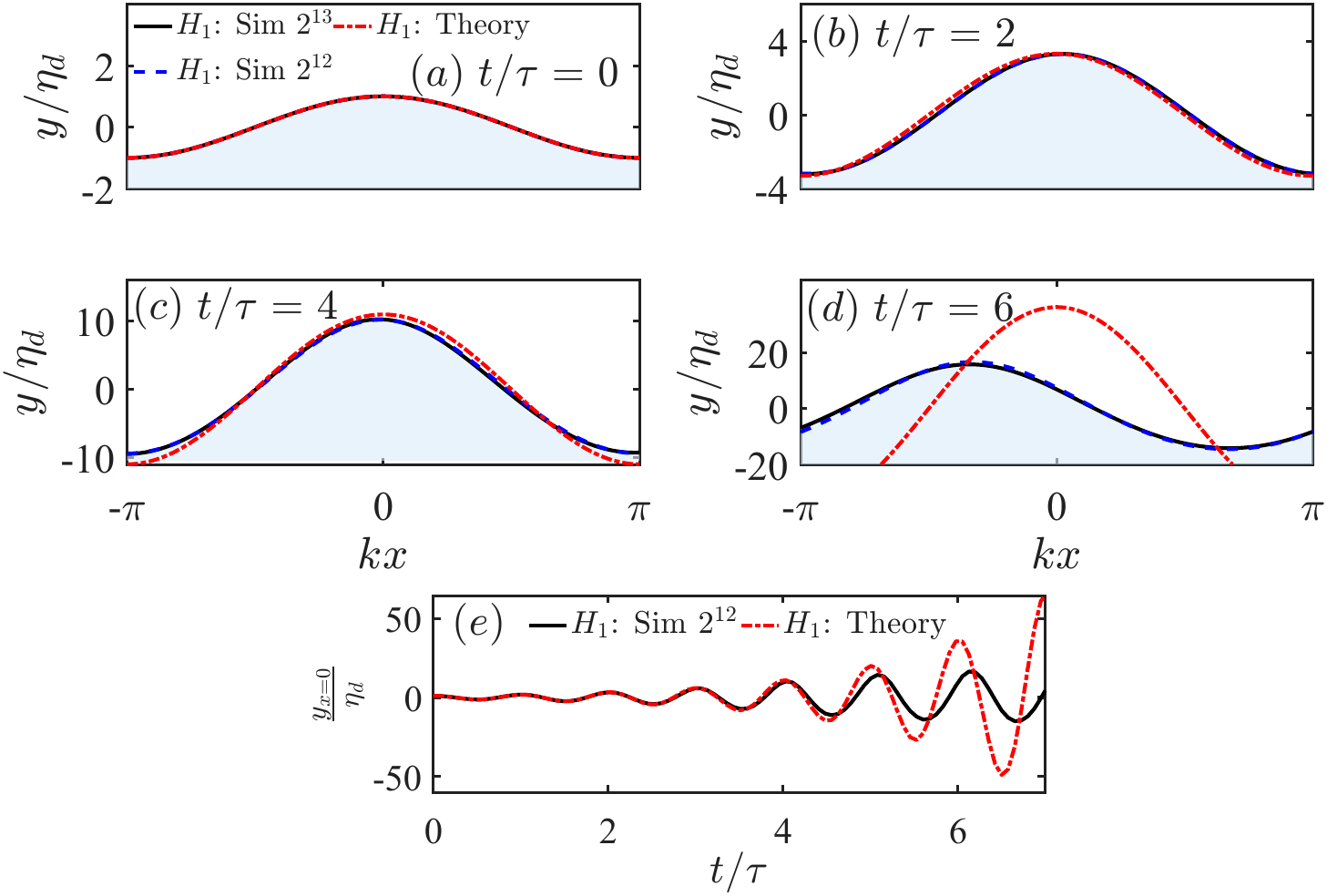}}
  \caption{Comparison of the density-vorticity interface ($H_1$) evolution between linear stability theory (dash-dotted red curves) and numerical simulations. (a-d) Spatial wave profiles at normalized times $t/\tau = 0, 2, 4$, and $6$, showing grid convergence between maximum refinement levels of 12 (dashed blue) and 13 (solid black). Time $t$ is normalized by the wave period $\tau$, defined as the time required for the most unstable eigenmode to travel one wavelength ($\lambda = \lambda^*/\Delta^* \approx 1.894$). (e) Temporal evolution of the interfacial displacement at the domain center ($x = 0$), comparing the theoretical exponential growth with the level-12 simulation (solid black). The vertical displacement is normalized by the initial amplitude $\eta_d$. Fixed parameters are $Fr = 0.8, Bo = 10^4$ and $r = 0.001$.}
 \label{fig9:LSA_basilisk_H1}
\end{figure}
Having discussed a spatially integrated measure of the surface wave characteristic (i.e. potential energy from eqn. \ref{eq:Energy}), we turn in this sub-section to describing the spatio-temporal features of the density-vorticity and the submerged vorticity interfaces for the fastest growing mode corresponding to a representative case with $Fr = 0.8$, $Bo = 10^4$, with $r = 0.001$. The corresponding dimensional parameters prescribed in the numerical solver are (shear layer thickness) $\Delta^* = 0.4\text{ cm}$ and $U^*_s = 15.839\text{ cm\,s}^{-1}$ (surface velocity). This yields a primary instability with a (fastest growing) dimensional wavenumber $k^* =8.292\text{ cm}^{-1}$ or wavelength $\lambda^* = 0.758\text{ cm}$, with growth rate $\sigma^* = k^*c_i^* = 1.864\text{ s}^{-1}$. These choice of parameters also reinforce the validity of the quasi-steady base state assumption. As established by \citet{morland1991waves}, the timescale for the growth of unstable modes ($1/\sigma^*$) must be vastly smaller than the viscous diffusion timescale ($\Delta^{*2}/\nu^*_w$), requiring $\sigma^* \Delta^{*2}/\nu^*_w \gg 1$. For our selected parameters above and kinematic viscosity for water (from table \ref{Tab1:physical_scales}(a)), this inequality is comfortably satisfied, yielding $\sigma^*\Delta^{*2}/\nu^*_w \approx 30$. Therefore, we can safely assume that the piecewise-linear background velocity profile remains frozen on the time-scale of growth of the linear instabilities. Additionally, these chosen parameter values ensure that the primary instability wavelength significantly exceeds the air-water capillary length, verifying that the system remains within the gravity-dominated regime (high Bond number). In non-dimensional form and at the aforementioned parameter values, the descriptors of the primary unstable wave are $(k, c_r, c_i, \lambda) = (3.317, 0.149, 0.014, 1.894)$ (refer to figure \ref{fig5:LSA_dispersion}), corresponding to a non-dimensional (real part) angular frequency of $\omega_r = k c_r = 0.494$. Figure \ref{fig9:LSA_basilisk_H1} compares the spatio-temporal evolution of the density-vorticity interface ($H_1$) against the analytical linear stability predictions. During the early stages of wave development ($t/\tau \le 2$, panel b), the simulated interfacial profile shows good agreement with the exponentially growing linear stability prediction. Incidentally, this also confirms that the tiny value of viscosity introduced in \S \ref{sec:Numerical_simulation} does not artificially damp the fundamental instability mechanism.

As the wave amplitude grows significantly beyond its initial value, the flow transitions into the nonlinear regime. By four wave cycles ($t/\tau = 4$, panel c), a clear divergence from the linear theory emerges, marking the onset of nonlinearity in its instantaneous shape. This is also mirrored in panel (e) in fig. \ref{fig9:LSA_basilisk_H1}, where the temporal signal for the density interface at the domain midpoint ($x=0$) accurately matches the theoretical exponential growth envelope up to roughly three wave periods before nonlinear effects restrict further exponential growth. Finally, these results also establish grid independence. Computations performed at maximum refinement levels of $2^{12}$ and $2^{13}$ show nearly overlapping spatial profiles throughout the linear and early nonlinear phases (panels a-c). Even deep into the nonlinear regime at $t/\tau = 6$ (panel d), the two resolutions match well, confirming that a resolution of $2^{12}$ is sufficient for capturing the nonlinear wave dynamics.  

\begin{figure}
  \centerline{\includegraphics[width = \textwidth]{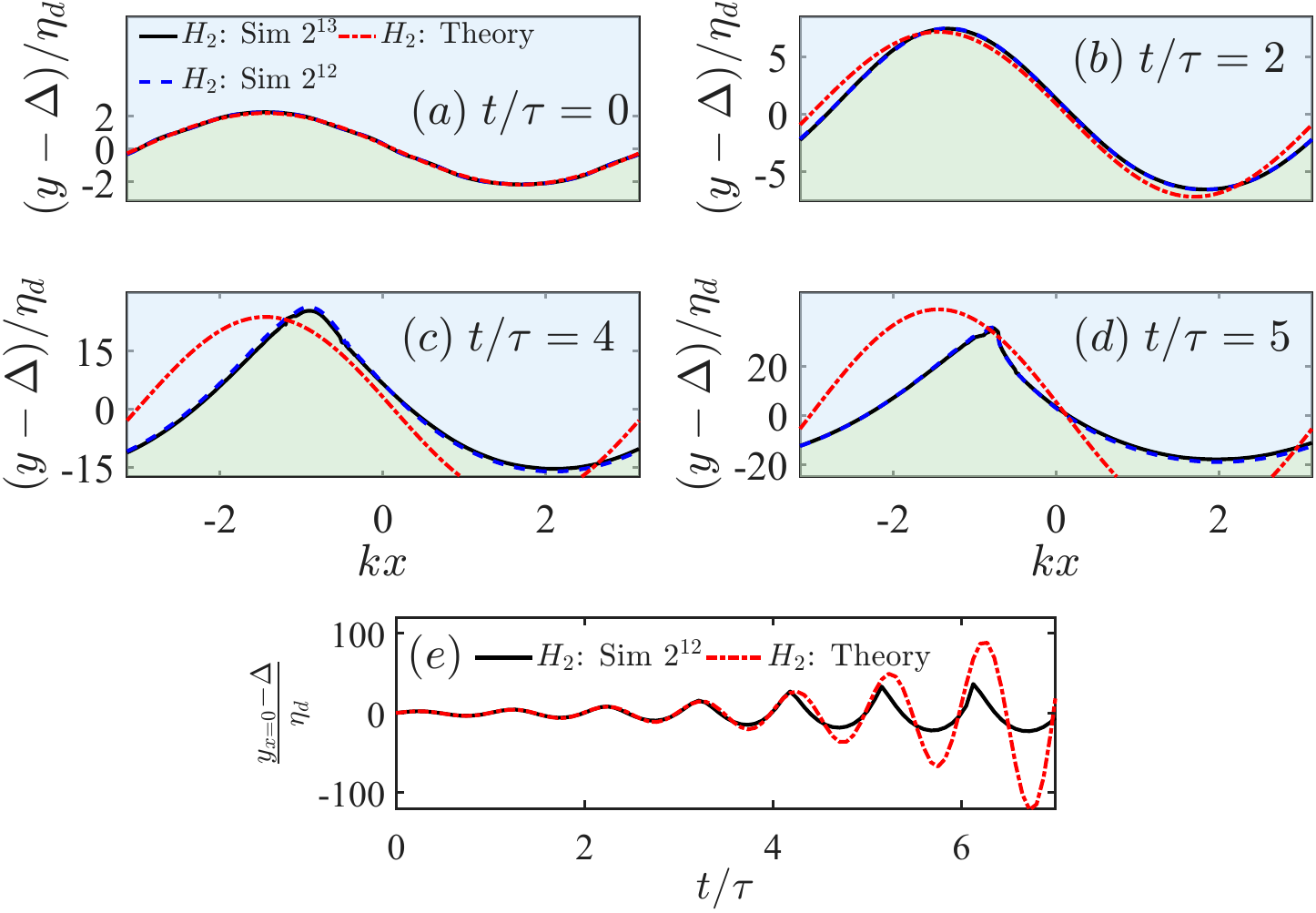}}
  \caption{Evolution of the subsurface vorticity interface ($H_2$) at $Fr = 0.8$. Panels (a-d) display the spatial wave profiles at consecutive normalized times $t/\tau = 0, 2, 4$, and $5$, and panel (e) presents the temporal displacement signal at the domain center ($x = 0$). Line styles, the wave period definition ($\tau$), and fixed physical parameters are identical to those detailed in figure \ref{fig9:LSA_basilisk_H1}. The vertical displacement is normalized by the initial $H_1$ amplitude $\eta_d$.}
 \label{fig10:LSA_basilisk_H2}
\end{figure}
Complementing the $H_1$ validation, fig. \ref{fig10:LSA_basilisk_H2} plots the concurrent evolution of the subsurface vorticity interface ($H_2$) from the same simulation as fig. \ref{fig9:LSA_basilisk_H1}. As shown in panels (a) and (b) of fig. \ref{fig10:LSA_basilisk_H2}, the solver accurately captures the linear growth phase up to $t/\tau = 2$ (a slight phase shift is observed). Because the subsurface vorticity wave initially possesses a larger amplitude\footnote{this can be shown easily in linear theory, see eqn. \ref{eq: LSA_coefficients} in appendix.} than the corresponding surface displacement ($H_1$), it shows a more rapid transition into the nonlinear regime. By $t/\tau = 4$ (panel c), and profoundly by $t/\tau = 5$ (panel d), the wave shape breaks linear symmetry evinced by a pronounced nonlinear steepening, characterized by crests sharpening and broadening of troughs - this is markedly more severe in $H_2$ than in $H_1$. The temporal signal in panel (e) not only validates the linear theory but also confirms that the exponential growth of the submerged vorticity interface is cut off shortly after three wave periods, similar to the upper density interface, as the system enters a nonlinear state. Consistent with the surface observations, the spatial profiles for both refinement levels ($2^{12}$ and $2^{13}$) remain nearly identical, proving that the grid adaptation on the velocity field adequately resolves the subsurface shear layer. We note that extracting the subsurface interface using the $-\Omega^*/2$ isocontour introduces small corrugations, observed in panels (b) and (c) of fig. \ref{fig10:LSA_basilisk_H2}. Because calculating vorticity requires numerical differentiation of the discrete velocity field, it is more susceptible to grid-scale noise than the volume-fraction tracking used for $H_1$.

\begin{figure}
	\centerline{\includegraphics[width = \textwidth]{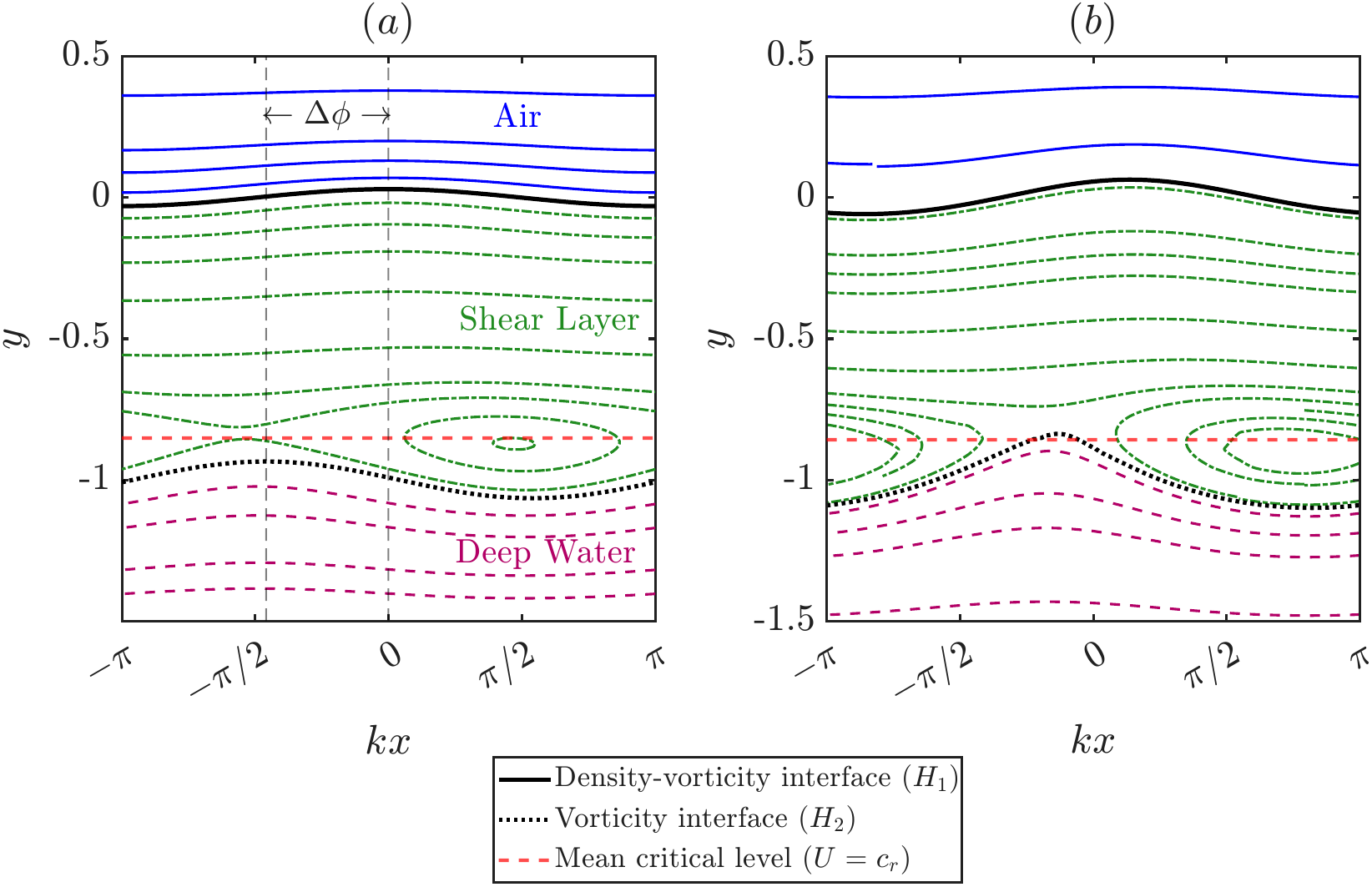}}
	\caption{Streamlines of (a) the most unstable linear eigenmode at $t=0$ and (b) the finite-amplitude wave at $t/\tau = 4.1$, viewed in a reference frame translating with the phase speed $c_r$ ($c_r = 0.149$ for panel a, and $c_r = 0.142$ for panel b). The local flow field across both panels is divided into the uniform air layer (solid blue curves), the subsurface water shear layer (dash-dotted green curves), and the deep irrotational water region (dashed magenta curves). In both regimes, the density-vorticity interface ($H_1$, solid black curve) and the subsurface vorticity interface ($H_2$, dotted black curve) exhibit a pronounced relative horizontal phase shift ($\Delta \phi$). The trapped Kelvin `cat's eye' vortical structure is clearly visible within the shear layer around the mean critical level ($U = c_r$, dashed red curve). Panel (b) confirms that this fundamental phase-locked topology predicted by linear stability theory is preserved in the fully nonlinear state. Fixed parameters are $Fr = 0.8$, $Bo = 10^4$, $r = 0.001$, and an initial linear steepness of $\eta_d k = 0.1$ (for panel (a)).}
	\label{fig11:basilisk_streamlines}
\end{figure}

\subsection{Streamline topology around the critical location}
To visualize the flow kinematics in the linear regime, figure \ref{fig11:basilisk_streamlines} presents the instantaneous streamlines at $t/\tau = 4.1$, computed in a reference frame translating with the phase speed $c_r$. For panel (a), $c_r = 0.149$ - the fastest growing mode phase speed from linear theory and for panel (b), $c_r = 0.142$ - from the same simulations as fig. \ref{fig9:LSA_basilisk_H1} and \ref{fig10:LSA_basilisk_H2}. Remarkably, the streamline topology of the fully nonlinear flow field (panel (b)) remains qualitatively consistent with the analytical predictions from linear theory (panel (a)). Plotted within the frames, the density-vorticity interface ($H_1$) and the subsurface vorticity interface ($H_2$) exhibit a pronounced relative horizontal phase shift ($\Delta \phi \neq 0$). Furthermore, the critical level ($U = c_r$) is confined within the subsurface shear layer in water. In the upper region of the domain, the local flow field in the air and upper shear layer remains strictly in phase with the perturbed density-vorticity interface ($H_1$). However, descending deeper into the shear layer, the streamlines undergo a gradual horizontal phase shift, culminating in a phase shift across the critical layer, characterized by the classic Kelvin `cat's eye' vortical structure. Below this critical location, the streamlines re-align to match the phase of the lower pure vorticity interface ($H_2$). Consequently, the critical location is seen to accommodate the horizontal phase shift ($\Delta \phi$) necessary in the streamline topology to account for the phase difference between the density-vorticity wave ($H_1$) and the submerged vorticity wave ($H_2$) predicted from linear theory. This phase-difference is crucial to the mechanism of the Holmboe instability.

\subsection{Nonlinear regime: Bound and free waves}\label{subsec:boundwaves}
\begin{figure}
  \centerline{\includegraphics[width = 0.8\textwidth]{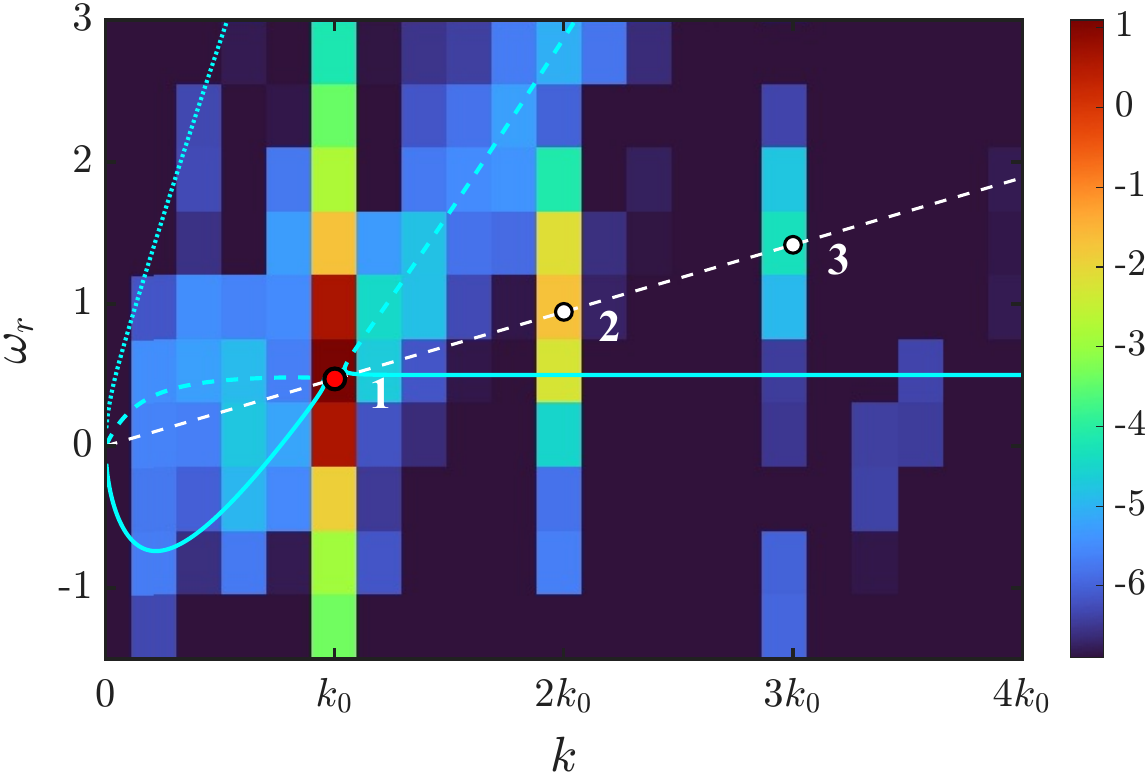}}
  \caption{Two-dimensional power spectral density, $S_{k\omega}=\log_{10}|\hat{\eta}_d(k,\omega_r)|^2$, of the density-vorticity interface ($H_1$) over the time window $3.5 \le t/\tau \le 4.5$ where $\hat{\eta}_d$ is the Fourier amplitude and $k,\omega_r$ are the non-dimensional spatial and temporal, angular frequencies respectively. The dominant primary wave $(k_0, \omega_{r,0})$ corresponding to the peak spectral energy is marked by the filled red circle. Higher-order bound harmonics are indicated by white circles; these phase-lock to the primary wave and consequently align along the white dashed line of constant phase speed ($c = \omega_r/k$). The analytical roots of the linear dispersion relation \eqref{eq: dispersion_relation_LSA_nondimensional} are overlaid as cyan curves (solid, dashed, and dotted). Physical parameters of the simulation correspond to figure \ref{fig9:LSA_basilisk_H1}. The range of unstable wavenumbers lie within a band around $k_0$, this being the fastest growing mode.}
 \label{fig12:spectral_analysis_H1}
\end{figure}
To investigate the composition of the nonlinear waves seen in the later stages of figs. \ref{fig9:LSA_basilisk_H1} and \ref{fig10:LSA_basilisk_H2}, we perform a spatio-temporal spectral analysis of the density-vorticity interface ($H_1$) over the time window $3.5 \le t/\tau \le 4.5$ in figure \ref{fig12:spectral_analysis_H1} (see caption for details). This time-interval corresponds to the ``\textit{nonlinear window}'' described earlier in \S\ref{subsec:energy}.
The spectral analysis computes the distribution of interfacial wave energy in the wavenumber (k) - frequency ($\omega_{r}$, subscript `r' indicating real part) domain through the two-dimensional power spectral density (PSD), $S_{k\omega}$ (see caption), from the raw spatio-temporal data for $H_1$ as obtained from simulations within the indicated time window. The spectra are computed using a windowed, spatio-temporal, Fast Fourier Transform (FFT) in MATLAB \citep{MATLAB:R2025a}. We isolate the peak spectral components and sort them in decreasing power. Full details regarding the spectral signal processing and peak extraction are provided in Appendix \ref{app:spectral_analysis}. To isolate physically meaningful harmonics from numerical noise, the spectral energy is truncated at a minimum threshold of $10^{-6}\times \max(S_{k\omega})$. The spectrum is dominated by the fundamental carrier wave at $(k_0, \omega_{r,0}) = (3.317, 0.472)$\footnote{we remind the reader that this is the fastest growing wavenumber for the linearised problem with the PL background state at the chosen parameters $Fr = 0.8$, $Bo = 10^4$, $r = 0.001$}. This numerically extracted frequency is lesser than the analytical linear stability result of $\omega_r = 0.494$ at the same wavenumber; the primary wave is thus seen to experience a small nonlinear frequency shift.

We recall that finite-amplitude nonlinear propagating waves, such as classical (unsheared) Stokes waves, are characterized by the presence of higher-order bound harmonics that travel at the exact phase speed of the primary carrier wave. To determine if the interfacial structure seen in our simulations indeed contain such phase-locked bound-wave components, we examine the PSD the for top three local energy maxima. Because these superharmonics are bound to the primary wave and travel at its phase-speed, they are constrained to lie along a line on the $k\text{-}\omega_r$ plane, as illustrated by the white dashed line in figure \ref{fig12:spectral_analysis_H1}. We note that the brightest spots (indicating the largest Fourier amplitude) for each of the wavenumbers $2k_0,3k_0$ in fig. \ref{fig12:spectral_analysis_H1}, lie on this straight line. The slope of this line defines the constant, non-dimensional phase speed of the fully nonlinear wave, yielding $c_{r,b} = \omega_r/k \approx 0.142$. This value is close to but slightly lesser than the analytically predicted linear phase speed of the fundamental carrier wave ($c_r = 0.149$). Table \ref{tab:Tab4_bound_harmonics_H1} summarizes the frequencies of the extracted spectral pairs and confirms the uniform phase speed across the fundamental mode and its bound harmonics for the density-vorticity wave ($H_1$).

\begin{table}
	\begin{center}
		\def~{\hphantom{0}}
		\begin{tabular}{lcccc}
			Mode  & $k$   &  $\omega_{r}$ & $S_{k\omega}$ & $c_{r}$ \\[3pt]
			1 Carrier mode ($M_1$)  & $3.317$ ($k_0$) &
			$0.472$ & $1.257e+01$ & $0.142$ \\
			2 Second bound harmonic ($M_2$)  & $6.634$ ($2k_0$) & $0.944$ & $2.031e-02$& $0.142$ \\
			3 Third bound harmonic ($M_3$) & $9.951$ ($3k_0$) & $1.417$ & $4.633e-05$ & $0.142$
		\end{tabular}
		\caption{Extracted non-dimensional wavenumbers, angular frequencies, power and phase speeds for top three physically distinct modes (local maxima) of the primary density-vorticity wave ($H_1$).}
		\label{tab:Tab4_bound_harmonics_H1}
	\end{center}
\end{table}

In addition to the bound superharmonics, the spectral decomposition in fig. \ref{fig12:spectral_analysis_H1}  also shows that the density-vorticity wave ($H_1$) also contains free wave components. Unlike the bound harmonics, these free components obey the linear dispersion relation \eqref{eq: dispersion_relation_LSA_nondimensional}.We observe that the energy of these free waves are concentrated around the dispersive branches associated with surface capillary-gravity modes (dashed and dotted cyan curves in \ref{fig12:spectral_analysis_H1}  coming from the roots of the dispersion relation). This is physically consistent, as these modes naturally govern the hydrodynamics of the upper interface (recalling the behavior established earlier in figure \ref{fig6:LSA_wave_interaction}. In contrast, the pure vorticity branch, represented by the solid horizontal cyan line at $\omega_r = 0.5$ (analogous to the magenta curve in figure \ref{fig6:LSA_wave_interaction}), exhibits negligible spectral energy. This is also expected, as the pure vorticity mode is localized within the subsurface shear layer ($H_2$) rather than the air-water interface. 

\begin{figure}
  \centerline{\includegraphics[width = 0.8\textwidth]{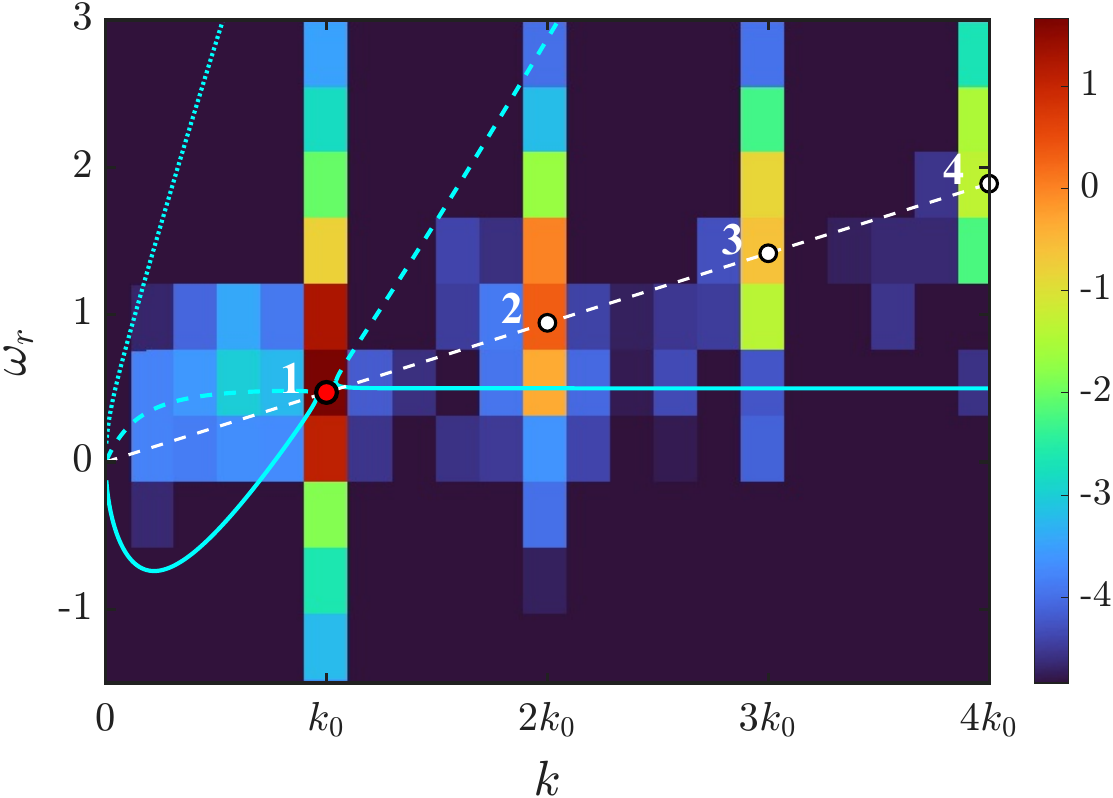}}
  \caption{Spectral analysis of the subsurface vorticity interface ($H_2$) during the nonlinear window ($3.5 \le t/\tau \le 4.5$). Two-dimensional power spectral density, $S_{k\omega} = \log_{10}|\hat{\eta}_v(k,\omega_r)|^2$, where $\hat{\eta}_v$ is the Fourier amplitude of the subsurface wave. Line styles, symbol definitions (identifying the primary wave, bound harmonics, and free components), and fixed physical parameters are identical to those detailed in figure \ref{fig12:spectral_analysis_H1}.}
 \label{fig13:spectral_analysis_H2}
\end{figure}
\begin{table}
	\begin{center}
		\def~{\hphantom{0}}
		\begin{tabular}{lcccc}
			Mode  & $k$   &  $\omega_{r}$ & $S_{k\omega}$ & $c_{r}$ \\[3pt]
			1 Carrier mode ($M_1$)  & $3.317$ ($k_0$) &
			$0.472$ & $4.586e+01$ & $0.142$ \\
			2 Second bound harmonic ($M_2$) & $6.634$ ($2k_0$) & $0.944$ & $2.114e+00$& $0.142$ \\
			3 Third bound harmonic ($M_3$) & $9.951$ ($3k_0$) & $1.417$ & $2.486e-01$ & $0.142$ \\
			4 Fourth bound harmonic ($M_4$) & $13.268$ ($4k_0$) & $1.889$ & $4.652e-02$ & $0.142$
		\end{tabular}
		\caption{Extracted non-dimensional wavenumbers, angular frequencies, power and phase speeds for top four physically distinct modes (local maxima) of the vorticity wave ($H_2$).}
		\label{tab:modes_H2}
	\end{center}
\end{table}
Figure \ref{fig13:spectral_analysis_H2} presents the spatio-temporal spectrum for the subsurface vorticity interface ($H_2$) under identical physical parameters and time window as figure \ref{fig12:spectral_analysis_H1}. The contours confirm that the subsurface wave also displays characteristics of a finite-amplitude permanent form progressive wave translating at a constant phase speed. This spectrum is also dominated by the fundamental carrier wave at $(k_0, \omega_{r,0}) = (3.317, 0.472)$. This extracted frequency is nearly identical to the primary frequency of the upper interface, confirming that the two waves are phase-locked. Consistent with observations from the density-vorticity interface, the local energy maxima of the $H_2$ spectrum in figure \ref{fig13:spectral_analysis_H2} identify the first three bound harmonics ($2k_0,3k_0$ and $4k_0$), see table \ref{tab:modes_H2} for details. As shown in the figure, these travel with the same constant phase speed as the carrier wave. Remarkably, the first three spectral pairs $(k, \omega)$ and the corresponding fully nonlinear phase speed ($c_{r} \approx 0.142$) are numerically identical upto three decimal places, to those reported for the upper interface in table \ref{tab:Tab4_bound_harmonics_H1}. One can therefore conclude that within this nonlinear window of observation, the two interfaces are not merely translating at the same phase-speed. Rather, their higher-order nonlinear bound-components are phase-locked with the carrier wave by translating at the same speed.

Turning to the free wave components in fig. \ref{fig13:spectral_analysis_H2}, we observe a fundamentally different result compared to the upper interface ($H_1$). The free waves here predominantly populate the pure vorticity branch (the horizontal solid cyan line at $\omega_r = 0.5$), reflecting the fact that this specific mode dictates the local hydrodynamics of the subsurface shear layer. Conversely, the surface capillary-gravity branches exhibit negligible spectral energy. In addition to this horizontal branch, we observe spectral energy distributed along the vertical axis at the fundamental wavenumber ($k_0$) across a range of frequencies. This vertical smearing also observed for other wavenumbers is a numerical artifact resulting from the finite size of the nonlinear time window chosen for the spectral analysis. Finally, figs. \ref{fig:H1_reconstruction} and \ref{fig:H2_reconstruction} in Appendix \ref{AppC:H2_H2_recon} show that reconstructing the interfaces $H_1$ and $H_2$ using only one free wave ($k_0$) and a few of its bound components as listed in Tables \ref{tab:Tab4_bound_harmonics_H1} and \ref{tab:modes_H2}, respectively, yields a strong match with the raw simulation waveforms. One notes that other free waves have not been used in this reconstruction.
 
The above analysis thus suggests that in the nonlinear time window of our observation, in a first approximation, the dominant part of the wave energy is concentrated in a progressive, finite-amplitude wave of permanent form arising in a sheared background flow. The free-wave component of this progressive wave has been chosen to be the fastest growing mode of the underlying linear instability with the same background shear. Simulations presented this far, strongly hint at the existence of such sheared, finite-amplitude progressive waves of permanent form. In \S\ref{sec:Connection to nonlinear waves} below, these finite-amplitude waves are computed from first principles and their form and speed are shown to agree with those seen in simulations.

\section{Nonlinear progressive waves of permanent form with background shear}\label{sec:Connection to nonlinear waves}

\subsection{Steady state formulation}\label{subsec:Nonlinear_formulation}
In this section, we derive the two-dimensional, finite-amplitude permanent-form shapes for both the density-vorticity wave ($H_1^*$) and the subsurface vorticity wave ($H_2^*$), as well as their speed for a given wavelength $\lambda^*$ and wave-height $h$ (see eqn. \ref{eq: wave height} for definition) corresponding to the system depicted in figure \ref{fig4:Schematic}. 
In what follows, we adopt the mathematical framework of \cite{milinazzo1990effect} and \cite{baumstein1999steady} towards this. However, our formulation critically incorporates the active air layer and a phase shift between the density and subsurface vorticity interfaces. Because this phase shift is the essential driver of the shear instability, we must express the wave profiles using both sine and cosine Fourier components (leading to an overdetermined problem, see numerical description later on). This distinguishes our work from \cite{milinazzo1990effect}, who suppressed this instability by assuming zero phase shift (cosine-only profiles). Furthermore, while \cite{baumstein1999steady} also retained the full Fourier series to model asymmetric waves, their analysis focused on bifurcations arising from multi-wavenumber interactions, rather than the primary shear-driven instability mechanism addressed here.
To achieve a steady flow, we depart from the laboratory reference frame used so far, transforming to a frame co-translating with the unknown constant wave phase speed, $c^*$. This is assumed to be parallel to the background velocity \citep{saffman1982finite}. In this co-moving frame, the unperturbed background velocity field becomes $\bU^*(y^*) - c^* \bex$ (see eq. \eqref{eq: base_velocity_PLP}). Consequently, the total steady stream function within each fluid layer ($j = a, w_1, w_2$) is defined as $\Psi_j^*(x^*,y^*) = \psi_j^*(x^*,y^*) - c^*y^*$.

Before formulating the nonlinear boundary value problem, it is necessary to redefine the characteristic scales. While the linear stability characteristics in the preceding sections were naturally scaled by the wind-speed and the shear-layer depth ($U^*_s, \Delta^*$), the finite-amplitude analysis is governed by the kinematics of the surface wave. Consequently, we adopt deep-water gravity wave scales, utilizing the inverse wavenumber $(1/k^*)$ and the linear gravity wave speed ($\sqrt{g^*/k^*}$) as the characteristic length and velocity scales, respectively. Table \ref{tab:Tab5_scales_FA} summarizes the complete set of characteristic scales chosen. We apply this non-dimensionalization and hereafter drop the asterisks and denote these new wave-scaled non-dimensional quantities with an over-tilde (e.g., $\tilde{\Psi}$).

\begin{table}
  \begin{center}
  \def~{\hphantom{0}}
  \begin{tabular}{lcl}
      Variable  & Characteristic scale   &   Description \\[3pt]
       Length   & $1/k^*$ & Inverse wavenumber \\
       Velocity   & $\sqrt{g^*/k^*}$ & Linear gravity wave speed \\
       Vorticity & $\sqrt{g^*k^*}$ & Wave-induced vorticity scale \\
       Stream function & $(1/k^*)\sqrt{g^*/k^*}$ & Kinematic flow scale \\
       Pressure & $\rho^*_w g^*/k^*$ & Convective pressure scale\\
       Surface tension & $\rho^*_wg^*/k^{*2}$ & Capillary force scale \\
       Density & $\rho^*_w$ & Quiescent water density \\
  \end{tabular}
  \caption{Characteristic scales for the pertinent variables for the finite-amplitude analysis.}
  \label{tab:Tab5_scales_FA}
  \end{center}
\end{table}

For the two-dimensional steady flow, the stream functions satisfy the following governing equations \citep{milinazzo1990effect}:
\begin{equation}
\left. \begin{array}{ll}
\displaystyle \tnabla^{2}\tPsi_a = 0,  \quad \quad \quad \text{for} \quad \tH_1(\tx) \leq \ty < \infty,\\[8pt]
\displaystyle \tnabla^{2}\tPsi_{w_1} = \tOmega, \quad \quad \text{for} \quad \tH_2(\tx) \leq \ty <  \tH_1(\tx),\\[8pt]
\displaystyle \tnabla^{2}\tPsi_{w_2} = 0, \quad \quad \quad \text{for} \quad -\infty < \ty < \tH_2(\tx). 
\end{array}\right\}
  \label{eq: governing_eqs_MCS_nondim}
\end{equation}
where $\tOmega$ is the non-dimensional form of same constant as in fig. \ref{fig4:Schematic}. Expressing the general boundary conditions \eqref{eq: BCs_dimensional} in terms of the steady relative stream functions, the governing equations \eqref{eq: governing_eqs_MCS_nondim} are subjected to the following boundary conditions (in non-dimensional form):
\begin{subequations}
    \begin{align}
       \tPsi_a &= 0, \quad &&\text{at} \quad \ty = \tH_1(\tx), \\
       \tPsi_{w_1} &= 0, \quad &&\text{at} \quad \ty = \tH_1(\tx),\\
       \tilde{P}_{w_1} - \tilde{P}_a &= \tT(\tilde{\bnabla}\bcdot\bn), \quad &&\text{at} \quad \ty = \tH_1(\tx),\\
       \tPsi_{w_1} &= K_1, \quad &&\text{at} \quad \ty = \tH_2(\tx),\\
       \tPsi_{w_1} &= \tPsi_{w_2}, \quad &&\text{at} \quad \ty = \tH_2(\tx),\\
       \frac{\partial \tPsi_{w_1}}{\partial \ty} &= \frac{\partial \tPsi_{w_2}}{\partial \ty}, \quad &&\text{at} \quad \ty = \tH_2(\tx).
    \end{align}
    \label{eq: nondim_BC_FA}
\end{subequations}
Conditions \eqref{eq: nondim_BC_FA}(\textit{a,b}) and (\textit{d,e}) enforce that both the density-vorticity ($\tH_1$) and subsurface vorticity ($\tH_2$) interfaces act as material streamlines, ensuring continuity across the respective boundaries. Here, the reference streamline is taken as zero at the upper interface, while $K_1$ denotes the constant streamline value at the lower interface. In \eqref{eq: nondim_BC_FA}(c), the curvature of the interface is taken as: $\tilde{\bnabla}\bcdot\bn = -\tilde{\mathcal{D}}^2\tH_1(\tx)(1+[\tilde{\mathcal{D}}\tH_1(\tx)]^2)^{-3/2}$, where $\tilde{\mathcal{D}} \equiv \mathrm{d}/\mathrm{d}\tx$.

Because the density-vorticity interface ($\tH_1$) acts as a steady material streamline, the Bernoulli equation is satisfied along this boundary in both the fluid phases. Consequently, we can evaluate the Bernoulli constants in the air and the subsurface shear layer as
\begin{equation*}
    \frac{1}{2}|\tilde{\bnabla}\tPsi_a|^2 + \frac{\tilde{P}_a}{r} + \tH_1 = b_1 \quad \text{and} \quad \frac{1}{2}|\tilde{\bnabla}\tPsi_{w_1}|^2 + \tilde{P}_{w_1} + \tH_1 = b_2,
\end{equation*}
where $b_1$ and $b_2$ are the non-dimensional Bernoulli constants in the air and water, respectively. By substituting the pressure terms into the dynamic boundary condition \eqref{eq: nondim_BC_FA}(\textit{c}), the normal stress balance is recast as:
\begin{equation}
    \frac{1}{2}|\tilde{\bnabla}\tPsi_{w_1}|^2 + \tH_1 - r \left( \frac{1}{2}|\tilde{\bnabla}\tPsi_a|^2 + \tH_1 \right) - \tT(\tilde{\bnabla}\bcdot\bn) = B_1,
    \label{eq: BC_norm_stress_FA_nondim}
\end{equation}
where $B_1 = b_2 - r\:b_1$ is the combined Bernoulli constant for the upper interface.

To capture the asymmetric structure of the finite-amplitude waves, specifically the horizontal phase shift between the upper and lower interfaces, the interfacial displacements and stream functions are expanded as full Fourier series truncated at $N$ modes. By explicitly incorporating both even (cosine) and odd (sine) harmonics, these expansions allow for a smooth phase variation from the air phase down into the deep water phase:
\begin{subequations}
    \begin{align}
        \tilde{H}_1(\tilde{x}) &= \sum_{n=1}^N \Big[\alpha_n \cos{(n\tilde{x})} + \hat{\alpha}_n \sin{(n\tilde{x})}\Big],\\
        \tilde{H}_2(\tilde{x}) &= -\tilde{\Delta} + \sum_{n=1}^N \Big[d_n \cos{(n\tilde{x})}+ \hat{d}_n \sin{(n\tilde{x})}\Big],\\
        \tilde{\Psi}_{a}(\tilde{x},\tilde{y}) &= (\tU_s - \tilde{c})\tilde{y} + Q_a + \sum_{n=1}^N e^{-n\tilde{y}}\Big[g_n  \cos{(n\tilde{x})}+ \hat{g}_n \sin{(n\tilde{x})}\Big],\\
        \tilde{\Psi}_{w_1}(\tilde{x},\tilde{y}) &= \frac{\tilde{\Omega}\tilde{y}^{2}}{2} - \hat{\tilde{c}}\tilde{y} + Q_{w_1} \nonumber \\ 
        &\quad + \sum_{n=1}^N \Big[\cos{(n\tilde{x})}\big\{a_n e^{n\tilde{y}} + b_n e^{-n(\tilde{y}+\tilde{\Delta})}\big\} + \sin{(n\tilde{x})}\big\{\hat{a}_n e^{n\tilde{y}} + \hat{b}_n e^{-n(\tilde{y}+\tilde{\Delta})}\big\}\Big],\\
        \tilde{\Psi}_{w_2}(\tilde{x},\tilde{y}) &= -\tilde{c}\tilde{y} + Q_{w_2} + \sum_{n=1}^N e^{n(\tilde{y}+\tilde{\Delta})}\Big[f_n \cos{(n\tilde{x})}+ \hat{f}_n \sin{(n\tilde{x})}\Big].
    \end{align}
    \label{eq: nondim_fourier_series}
\end{subequations}
where $\hat{\tilde{c}} = \tilde{c} - \tU_s = \tilde{c} - \tilde{\Omega}\tilde{\Delta}$, and $Q_a$, $Q_{w_1}$, and $Q_{w_2}$ represent the arbitrary constants of integration for the base stream functions. 

Out task now is to compute the unknown Fourier coefficients and the nonlinear phase speed ($\tilde{c}$) in eqns. \ref{eq: nondim_fourier_series}. For this, the truncated series \eqref{eq: nondim_fourier_series} are substituted into the boundary conditions \eqref{eq: nondim_BC_FA} and \eqref{eq: BC_norm_stress_FA_nondim}, yielding a system of nonlinear algebraic equations. In addition to the six boundary conditions, we need two more equations for enforcing wave height and for fixing the zero of the horizontal ($\tilde{x}$) direction. The first requirement leads to
\begin{equation}
    \max_{\tilde{x}} [\tilde{H}_1(\tilde{x})] - \min_{\tilde{x}} [\tilde{H}_1(\tilde{x})] = h, \label{eq: wave height}
\end{equation}
where $h$ is the prescribed (non-dimensional) peak-to-trough wave height of the density-vorticity interface. The maxima and minima in eqn. \ref{eq: wave height} of $\tilde{H}_1$ are determining by identifying their approximate locations on the discrete grid and refining them using a root-finding algorithm (e.g., \texttt{fzero} in MATLAB \citep{MATLAB:R2025a}) applied to the analytical spatial derivative of the interface. 
Secondly, by choosing a zero-slope extremum (crest or trough) as the origin $\tilde{x}=0$, we have
\begin{equation}
    \left. \frac{d\tilde{H}_1}{d\tilde{x}} \right|_{\tilde{x}=0} = 0, 
\end{equation}
which, based on the spatial expansion \eqref{eq: nondim_fourier_series}(\textit{a}), analytically reduces to:
\begin{equation}
    \sum_{n=1}^N n\: \hat{\alpha}_n  = 0. \label{eq: phase_pinning}
\end{equation}
We note that the wave-based non-dimensional numbers appearing in the above equations are related to dimensional numbers in table \ref{tab:Tab3_nondim_numbers_LSA} as: $\tilde{\Delta} = k$, $\tU_s = Fr\sqrt{k}$, $\tilde{\Omega} = Fr/\sqrt{k}$ and $\tilde{T} = k^2/Bo$. For a given set of fixed physical parameters ($Fr$, $Bo$, $r$ and $h$ - fixing all other non-dimensional parameters), the system comprises $12N+7$ unknowns:
the wave phase speeds ($\tilde{c}$, $\hat{\tilde{c}}$), the integration constants ($Q_a, Q_{w_1}, Q_{w_2}, B_1, K_1$), and the $12N$ Fourier coefficients. Notably, while $\hat{\tilde{c}}$ is physically constrained by the base flow, it is deliberately decoupled and treated as an independent unknown during the root-finding process to serve as a post-convergence consistency check (ensuring $\hat{\tilde{c}} = \tilde{c} - \tU_s$). The six continuous boundary conditions are discretized by evaluating them at $2N+1$ equally spaced collocation points along the horizontal axis from $\tx = 0$ to $2\pi$ ($\tilde{x}_i = 2\pi (i-1) / (2N+1)$ for $i = 1, 2, \dots, 2N+1$). Including the wave height and phase-pinning constraints, this yields an expanded system of $12N+8$ algebraic residual equations (detailed explicitly in Appendix \ref{app:residual_equations}). To accommodate this overdetermined system (number of equations is one more than number of unknowns), the problem is solved iteratively utilizing the Levenberg-Marquardt algorithm (implemented with MATLAB's \texttt{fsolve} \cite{MATLAB:R2025a}). The solver is supplied with the exact analytical Jacobian matrix of the residual equations. This least-squares minimization approach inverts the system, employing the linear stability eigenmodes detailed in Appendix \ref{app:initial_guess_LSA} as the initial guess to initiate the numerical continuation strategy outlined below.

\subsection{Numerical continuation}\label{subsec:continuation}

To continue the linearly unstable eigenmodes obtained theoretically to the finite-amplitude propagating states observed in simulations in \S \ref{sec:Numerical_simulation}, we formulate a two-stage numerical continuation pathway. From the reference numerical simulations ($Fr=0.8$, $Bo=10^4$, and $r=0.001$), we extract a finite-amplitude state within the nonlinear window characterized by a peak-to-trough wave height $h = h_{target}$. Under the wave-based non-dimensionalisation adopted in this section (lengths normalized by $1/k^*$) and for aforementioned parameters, the non-dimensional shear depth of this primary (fastest growing) unstable mode is $\tilde{\Delta} = k^*\Delta^* = 3.317$. To prevent divergence, the solver is however initiated at a reduced shear depth $\tilde{\Delta} = 3.172$. For $Fr=0.8$, this physically corresponds to a wavenumber which is lower than that of the fastest growing mode ($\tilde{\Delta} = 3.171$), see fig. \ref{fig3:Regime_map_unstable}. Primary continuation is then performed with the linearised eigenmode as initial guess (initial $h=10^{-3}$) and by incrementally stepping the wave height until the target amplitude ($h = h_{target}$) is achieved. 
With the wave height locked at $h_{target}$, a secondary continuation is then performed increasing the shear depth parameter $\tilde{\Delta}$ incrementally from $3.172$ up to the physical target value of $3.317$. Throughout this iterative root-finding process, the Fourier series in \eqref{eq: nondim_fourier_series} are truncated at $N = 40$ modes, this turns out to be sufficient spectral resolution to accurately resolve the severe nonlinear steepening of the interfaces.

The above framework has been rigorously validated in both the irrotational and rotational flow limits. As detailed in Appendix \ref{appC:validation_FA}, our formulation demonstrates excellent quantitative agreement with the benchmark studies of \citet{saffman1982finite} and \citet{milinazzo1990effect}, who primarily reported permanent-form shapes corresponding to eigenmodes which are linearly stable under background shear. It is important to mention that at sufficiently large wave heights ($h$), the physical-space collocation solver experiences convergence issues.  Consequently, we do not report steady-state solutions for large $h$. Additionally, although the numerical continuation is initialized with the finite phase difference between $H_1$ and $H_2$ predicted by linear theory, this phase information is not preserved during the procedure; this is a notable limitation of our current continuation strategy. To resolve limiting finite-amplitude waves with background shear with even more accuracy, presumably a more sophisticated numerical framework is required, such as the conformal mapping techniques employed by researchers \citep{saffman1982finite,dyachenko2019stokesa,dyachenko2019stokes,murashige2020stability} and this is deferred for subsequent work. In the following sub-section, we present the finite-amplitude solutions that have been obtained using the continuation strategy elucidated here.

\subsection{Finite-amplitude permanent form progressive waves}\label{subsec:FA_permanent_form_sol}
\begin{figure}
	\centerline{\includegraphics[width = \textwidth]{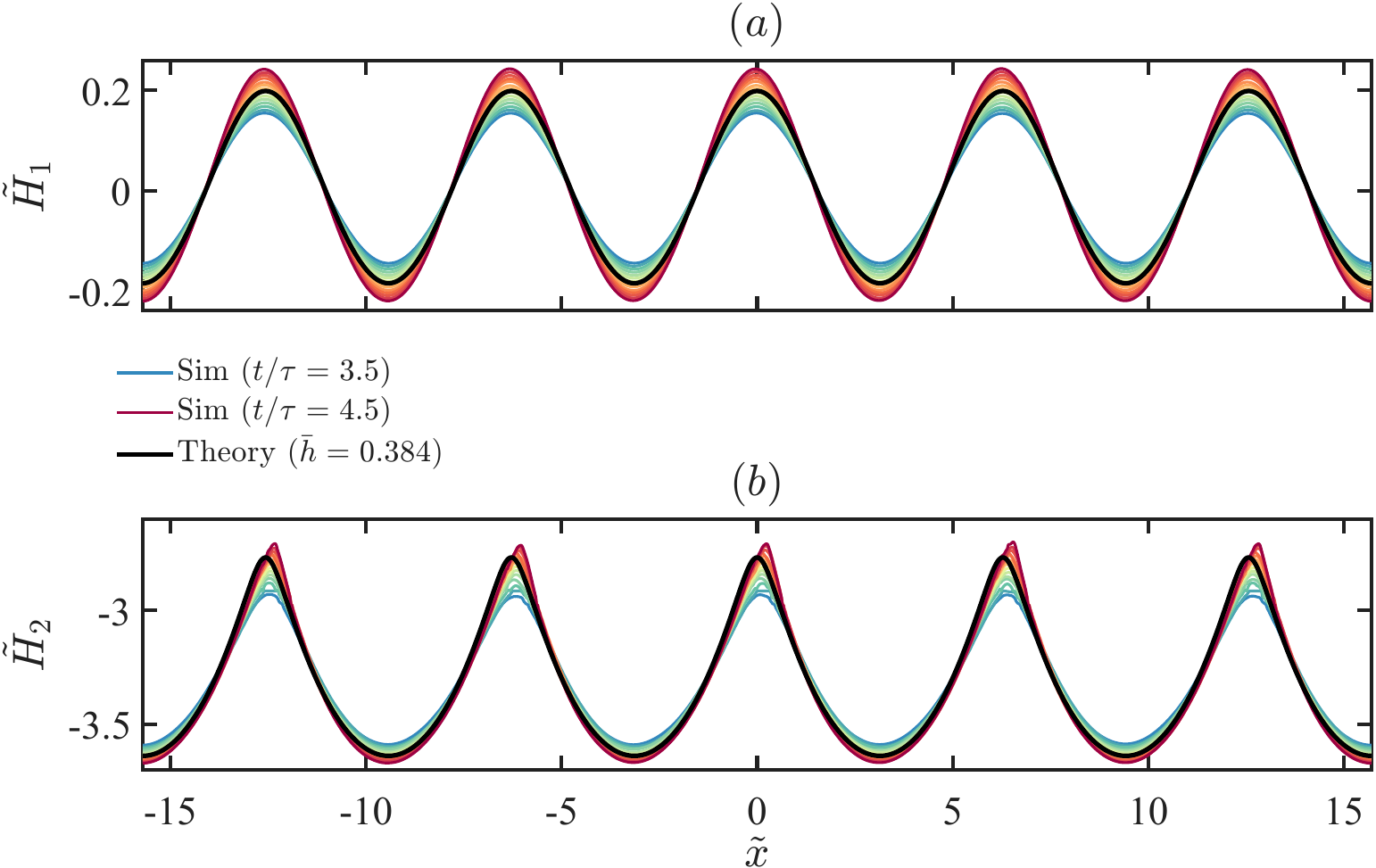}}
	\caption{Spatial profiles of (a) the density-vorticity interface ($\tH_1$) and (b) the subsurface vorticity interface ($\tH_2$). Solid colored lines represent transient wave profiles extracted from the multiphase numerical simulations over the nonlinear window $t/\tau \in [3.5, 4.5]$. To account for translation of the wave in the simulations, the simulated profiles ($\tilde{x}$) are shifted to align their wave crests at the origin ($\tilde{x}=0$). The thick solid black lines denote the exact permanent-form wave (as obtained from numerical continuation) evaluated at the time-averaged density-vorticity wave height ($\bar{h} = 0.384$). System parameters: $Fr = 0.8$, $Bo = 10^4$, $r = 0.001$, and $\tilde{\Delta} = 3.317$.}
	\label{fig14:shape_match}
\end{figure}
\begin{figure}
	\centerline{\includegraphics[width = \textwidth]{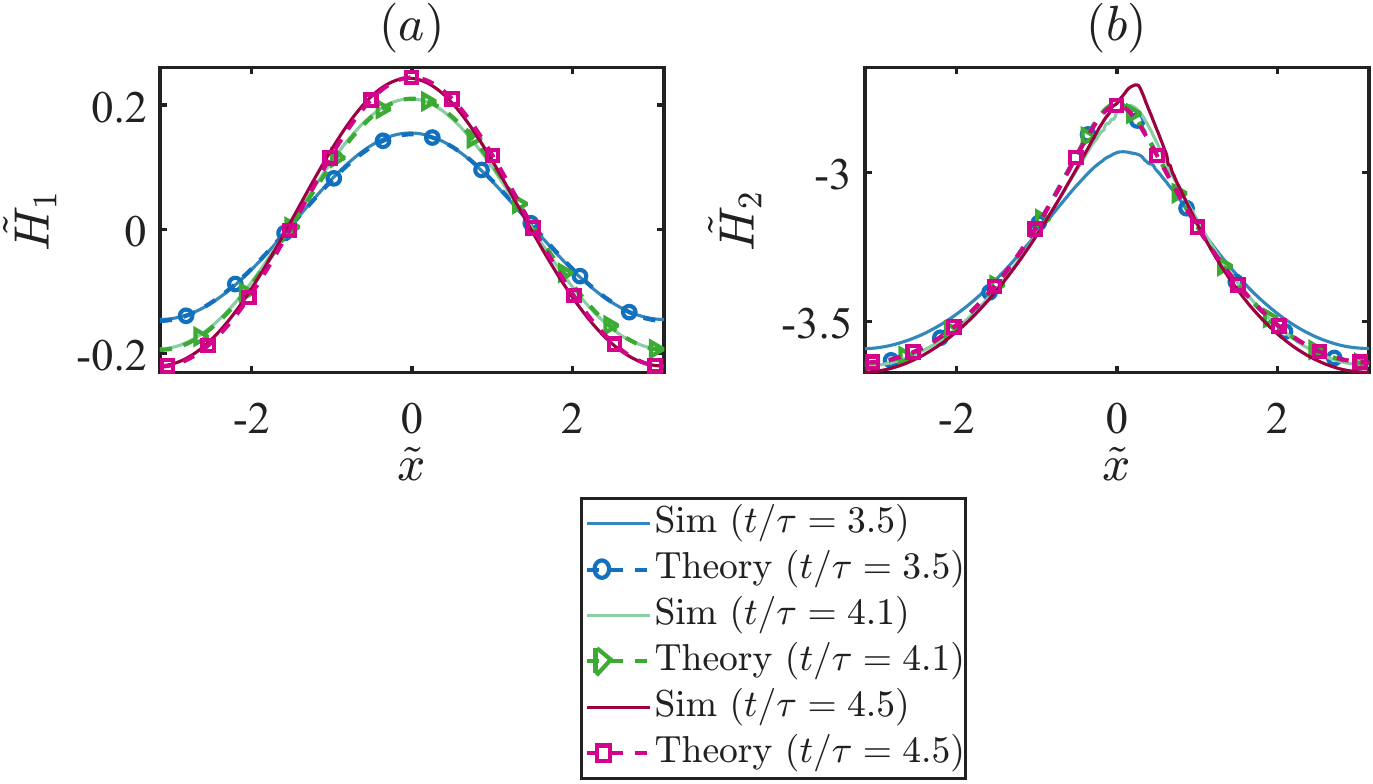}}
	\caption{Comparison of instantaneous simulated interfacial profiles (solid curves) against exact finite-amplitude steady-state theoretical solutions (dashed curves with symbols) at individual times during the nonlinear window for $Fr = 0.8$.
		(a) The non-dimensional density interface ($\tilde{H}_1$) exhibits excellent agreement with the theory across all sampled times ($t/\tau = 3.5, 4.1$, and $4.5$).
		(b) The subsurface vorticity interface ($\tilde{H}_2$) demonstrates strong match with the exact theory at $t/\tau = 4.1$. At $t/\tau = 3.5$, the steady-state theory overpredicts the simulated wave amplitude, while at $t/\tau = 4.5$, the simulated profile deviates from the theory (see the rightward bending of the crest) due to the onset of vorticity wave breaking.}
	\label{fig15:individual_shape_match}
\end{figure}
Figures \ref{fig14:shape_match}(a),(b) compare the exact steady-state theoretical solutions (black) against wave profiles extracted from nonlinear simulations for several time instants within the nonlinear time window ($t/\tau \in [3.5, 4.5]$). Because the simulated waves exhibit slow residual growth and slight phase drift during this evolution, the profiles have been shifted to align their crests at origin for comparison against their theoretically obtained counterparts. In figure \ref{fig14:shape_match}(a), for the primary density-vorticity interface ($\tilde{H}_1$), the theoretically obtained waveprofile (from numerical continuation) evaluated at the time-averaged wave height ($\bar{h} = 0.384$) is seen to perfectly lie midway of the transient numerical envelope. Figure \ref{fig15:individual_shape_match}(a) presents one-to-one comparison between analytical profiles and simulation profiles for $\tilde{H}_1$ at various time instants. This figure \ref{fig15:individual_shape_match}(a), demonstrates that a match holds continuously for $\tilde{H}_1$ as the wave height grows from $h \approx 0.3$ at $t/\tau = 3.5$ to $h \approx 0.465$ at $t/\tau = 4.5$.

The subsurface vorticity interface ($\tilde{H}_2$), on the other hand, maintains a significantly larger time-averaged height ($\bar{h}_v = 0.832$) as seen in fig. \ref{fig14:shape_match}(b). While the time-averaged theoretical profile captures essential aspects of its nonlinear shape (figure \ref{fig14:shape_match}(b)), the instantaneous snapshots (figure \ref{fig15:individual_shape_match}(b)) reveal key differences. Theory overpredicts the wave amplitude at $t/\tau = 3.5$, aligns optimally at $t/\tau = 4.1$, and again diverges from simulations at $t/\tau = 4.5$ as the simulated crest is higher and bends rightward. This localized deformation marks the onset of subsurface wave breaking in simulations. Finally, at $\bar{h} = 0.384$ (time averaged value of density-vorticity wave height in fig. \ref{fig14:shape_match}(a)), the steady-state solver predicts a wave-scaled non-dimensional phase speed of $\tilde{c} = 0.215$. Converting this to the base-flow scaling (\S \ref{sec:Linear_stability}) yields $c = \tilde{c} Fr^{-1}\tilde{\Delta}^{-1/2} \approx 0.148$. This theoretically predicted value is in excellent agreement with the fully nonlinear phase speed extracted through spatio-temporal spectral analysis ($c_{r} = 0.142$, see table \ref{tab:Tab4_bound_harmonics_H1}) of simulation data. The small relative error of approximately 4\%, along with the shape match confirms that the finite-amplitude theoretical framework accurately predicts the evolving wave dynamics, in the nonlinear time window.

\section{Conclusions}\label{sec:Conclusions}

In this work, we have investigated the linear and fully nonlinear evolution of surface waves arising from an inviscid, shear-driven instability at low Froude numbers in the gravity-dominated ($Bo >> 1$) regime; this configuration is representative of wind-drift instability generated surface waves in the ocean. The primary wave-generation occurs through linear instability of the sheared background state. This comprises a uniform atmospheric current in air over a constant-vorticity drift layer in water of finite thickness overlying deep, irrotational water, along with a flat air-water density interface. A classical normal-mode temporal linear stability analysis of this background state isolates the fastest growing unstable eigenmodes. Using wave interactions and at the Froude number range of current interest, we demonstrate that this instability is of the Holmboe type, arising above a critical Froude number due to phase-locking and mutual reinforcement of the density-vorticity and the submerged vorticity waves. The fastest growing eigenmodes seed two-phase Navier-Stokes simulations, capturing the transient evolution of the instability - from the initial exponential growth to the development of nonlinear propagating density and vorticity waveforms which can even break. To isolate the nonlinear propagating wave of permanent form contained within the simulation waveform, we formulate an exact, fully nonlinear boundary-value problem which is then solved numerically. 

Our transient simulations verify that the primary unstable mode exhibits an initial exponential growth, mirroring linearised predictions accurately. This exponential growth is eventually cutoff culminating in a quasi-saturated nonlinear state. Within the time window where this state is observed, the density-vorticity wave at the air-water interface and the subsurface vorticity wave, phase-lock into a coherent travelling structure. This double-interface system translates at a near constant phase speed, while undergoing marginal deformation due to the presence of other free-wave components, in addition to also displaying slow residual growth. We hypothesize that this secondary growth could potentially originate from linear instability of the sheared finite-amplitude state, much as subharmonic instability of the unsheared Stokes wave can lead to breaking \citep{iafrati2013modulational}. We also observe that breaking of the submerged vorticity interface (time instants corresponding to blue arrows in fig. \ref{fig8:Wave_energy}) significantly precedes that of the density-vorticity interface (time instants for the last data points for $Fr=0.8,1.0$ in fig. \ref{fig8:Wave_energy}). Incidentally, in simulations the density-vorticity interface when it breaks, tends to do so backwards (image not shown) i.e. in the opposite direction of the background shear in water. While such backward breaking is well documented for high-density-ratio interfaces (see fig. 2 of \cite{hult2009breaking}), it also emerges here in a low-density-ratio air-water system with a background wind-drift layer. 

We recall that the experimental results of \citet{lake1978new} highlighted the appearance of permanent-form progressive waves in their wind-wave experiments, although background shear (presumably primarily from air in \cite{lake1978new}) was not explicitly accounted for, in the description of their Stokes waves. Our spatio-temporal spectral analysis of the simulated wave field explicitly confirms this mechanism, specifically when the background shear originates from water instead of air. Furthermore, the exact travelling-wave solutions derived independently from our boundary-value formulation extending the framework of \cite{milinazzo1990effect}, exhibit very good agreement with the time-averaged wave profiles from the  Navier-Stokes simulations within the nonlinear time window. This agreement highlights that dynamically evolving interfacial waves, born from infinitesimal amplitude unstable sheared eigenmodes, can potentially get ``attracted'' toward these exact, sheared, Stokes-like propagating waves of permanent form, when the latter are linearly stable. Demonstrating this theoretically and from first principles is a challenging analytical exercise requiring accounting for viscosity within a weakly nonlinear formulation. Additionally, linear (in)stability characteristics of the finite-amplitude propagating states obtained here, require investigation and will be reported subsequently. We conclude by noting the recent interesting study by \cite{mogilevskiy2026emerging} who investigate a modified, linearised version of the Ginsburg-Landau equation in an IBVP formulation for wind-waves. \\

\noindent{\bf Declaration of Interests\bf{.}} The authors report no conflict of interest.
    
\section*{Acknowledgments:}
	We acknowledge support from DST-SERB (Govt. India)
	grants MTR/2019/001240, \\CRG/2020/003707, SPR/2021/000536 and Ministry of Education grant \\MoE-STARS/STARS2/2023-0595 on topics related to wind-wave generation and wave-breaking. The tenure of SNJ, supported by an institute postdoctoral fellowship (IPDF) from IIT Bombay is thankfully acknowledged.

\appendix
\normalsize
\section{Linear stability  - details and validation}\label{appA:linear_stability}
\subsection{Complex linear stability coefficients}
\label{appA:complex_coefficients_LSA}
From \eqref{eq: BCs_LSA_nondim}, the linear stability variables $\eta_v,A,B,D,E$ can be written in terms of $\eta_d$ (free variable), $c,k$ as: 
\begin{subequations}
    \begin{align}
        \eta_v &= \eta_d (c - 1) \frac{2 k e^{-k}}{2 c k + e^{-2 k} - 1}, \quad A = \eta_d(c-1), \\
        B &= \frac{\eta_d(c-1)(2ck-1)}{2 c k + e^{-2 k} - 1}, \quad D = \frac{\eta_d(c-1)e^{-2 k}}{2 c k + e^{-2 k} - 1}, \quad E = \frac{\eta_d(c-1)(2ck)}{2 c k + e^{-2 k} - 1}.
    \end{align}
    \label{eq: LSA_coefficients}
\end{subequations}

\subsection{Validation of the linear stability analysis}
\label{appA:validation_LSA}

\begin{figure}[ht!]
  \centering
  \includegraphics[width = \textwidth]{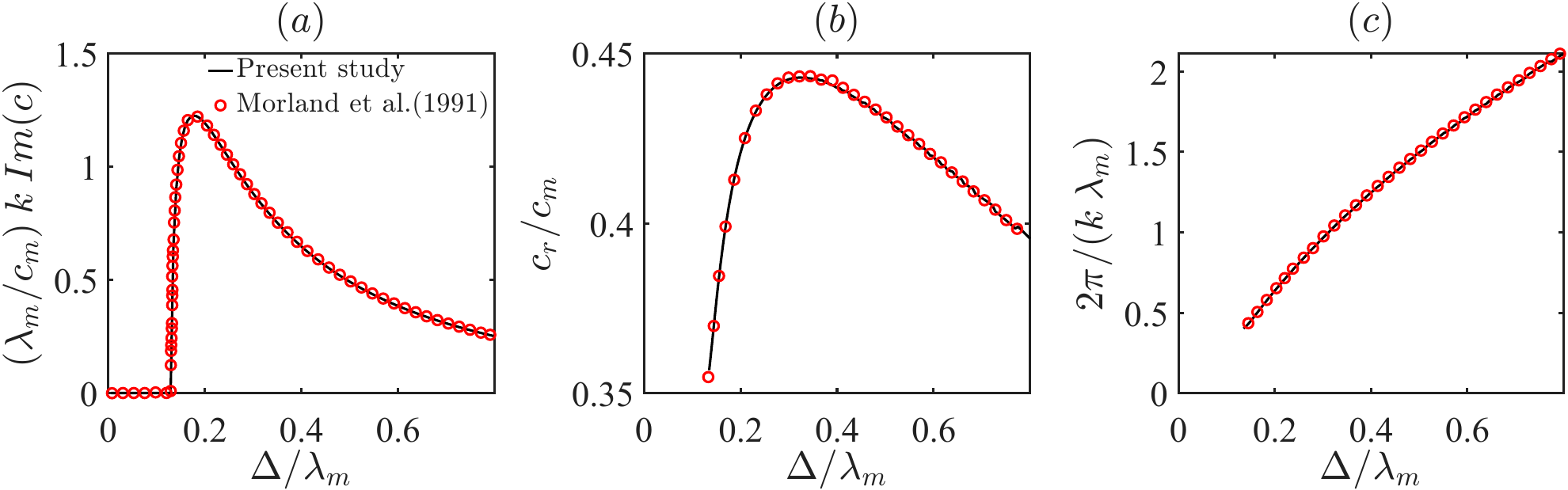}
  \caption{Properties of the most unstable mode as a function of normalized vorticity layer thickness for $U^*_s = 2c^*_m$. (a) Normalized growth rate; (b) normalized phase speed; (c) normalized wavelength. The solid curves are our results and the symbols represent the data reported by \citet{morland1991waves}.} 
\label{fig:validation_morland}
\end{figure}

Figure \ref{fig:validation_morland} compares our dispersion relation results from \eqref{eq: dispersion_relation_LSA_nondimensional} (solid lines) against the study by \citet{morland1991waves} (symbols). The panels depict the properties of the most unstable mode as a function of the non-dimensional vorticity layer thickness, $\Delta^*/\lambda^*_m$, for the piecewise-linear profile with $U^*_s = 2c^*_m$. Here, $c^*_m = (4g^*T^*)^{1/4}$ and $\lambda^*_m = 2\pi(T^*/g^*)^{1/2}$ are the phase speed and wavelength of the slowest capillary-gravity wave, respectively. The properties shown are: (a) non-dimensional growth rate, $(\lambda^*_m/c^*_m)k^*\mathrm{Im}(c^*)$; (b) non-dimensional phase speed, $\mathrm{Re}(c^*)/c^*_m$; and (c) non-dimensional wavelength, $2\pi/(k^*\lambda^*_m)$. The excellent agreement between the two studies validates our numerical implementation of the linear stability analysis.

\section{Numerical simulation details}
\subsection{Justification of the computational domain size}\label{appB:domain_size}

To guarantee domain size independence of simulation results, the required vertical extent of the computational domain, $L$, is determined analytically from the far-field exponential decay of the computed finite-amplitude wave in \S \ref{sec:Connection to nonlinear waves} for $Fr = 0.8$, at the average wave-height of $\bar{h} = 0.384$. Deep in the stationary water layer ($y^* \to -\infty$), higher nonlinear harmonics vanish, and the horizontal perturbation velocity is governed exclusively by the fundamental mode:
\begin{equation}
	\max |u_{w,2}^*(y^*)| \approx k^* \sqrt{(f_1^*)^2 + (\hat{f}_1^*)^2} \, \mathrm{e}^{k^*(y^*+\Delta^*)},
\end{equation}
where $f_1^*$ and $\hat{f}_1^*$ are the dimensional stream function amplitudes of the fundamental mode, converted from the non-dimensional collocation solution (\S \ref{subsec:Nonlinear_formulation}) using the characteristic scale $\psi_c = (k^*)^{-1}\sqrt{g^*/k^*}$. The effectively infinite depth is defined as the coordinate $y^*$ where this perturbation velocity drops below a negligible numerical threshold, $\epsilon U^*_s$ (with $\epsilon = 10^{-6}$). Solving for this required depth yields:
\begin{equation}
	y^*_w = -\Delta^* + \frac{1}{k^*} \ln \left( \frac{\epsilon U^*_s}{k^* \sqrt{(f_1^*)^2 + (\hat{f}_1^*)^2}} \right).
\end{equation}
For the parameters investigated herein ($k^* = 8.292 \text{ cm}^{-1}$, $\Delta^* = 0.4 \text{ cm}$, $U^*_s = 15.839 \text{ cm}\,\text{s}^{-1}$) and the computed fundamental amplitudes $f_1^* \approx 0.098 \text{ cm}^2\,\text{s}^{-1}$ and $\hat{f}_1^* \approx 4.213 \times 10^{-9} \text{ cm}^2\,\text{s}^{-1}$ (machine zero), the velocity decays to the $10^{-6} U^*_s$ threshold at a depth of $y^*_w \approx -1.708 \text{ cm}$.
Furthermore, evaluating the domain extent based solely on the horizontal perturbation velocity is mathematically exhaustive for deep-water irrotational flow. For the fundamental harmonic, the vertical velocity $v^* = -\partial\psi^*/\partial x^*$ possesses an identical amplitude envelope to the horizontal velocity $u^* = \partial\psi^*/\partial y^*$. Bounding the horizontal velocity component implicitly bounds the total perturbation vector magnitude. Because the required depth scales logarithmically with the wave amplitude, even the maximum transient peak amplitudes observed during early-time wave evolution require negligible additional vertical space. Consequently, the domain size evaluated from the saturated finite-amplitude form acts as a conservative, uniformly valid spatial estimate for the entire time-evolution of the flow.

To confirm that the upper boundary is equally unrestrictive, an analogous estimate is constructed for the uniform air domain. The fundamental mode of the perturbation velocity in the air decays as $\mathrm{e}^{-k^*y^*}$, yielding a maximum velocity envelope of:
\begin{equation}
	|u_{pert,a}^*| \approx k^* |g_1^*| \, \mathrm{e}^{-k^*y^*},
\end{equation}
where $g_1^*$ is the dimensional stream function amplitude of the fundamental mode in the air. Applying the characteristic scaling to the collocation output yields $|g_1^*| \approx 0.316 \text{ cm}^2\,\text{s}^{-1}$. Equating this envelope to the threshold $\epsilon U^*_s$ yields the required upper domain height:
\begin{equation}
	y^*_a = -\frac{1}{k^*} \ln \left( \frac{\epsilon U^*_s}{k^* |g_1^*|} \right) \approx 1.449 \text{ cm}.
\end{equation}
Because the denser water phase contains the rotational shear layer, the hydrodynamics penetrate significantly deeper into the water ($-1.708 \text{ cm}$) than into the overlying air ($+1.449 \text{ cm}$). The selected symmetric computational domain spans $y^* \in [-L/2, L/2]$. By setting $L = 5\lambda^*$ ($3.789 \text{ cm}$), the physical boundaries are placed at $y^* = \pm 1.894 \text{ cm}$, safely exceeding the required decay distances in both fluid phases (i.e. $L/2 > \max{(|y^*_w|, y^*_a)}$) and guaranteeing an effectively infinite domain.

\subsection{Spectral analysis methodology}\label{app:spectral_analysis}

To analyze the spatio-temporal energy distribution of the simulated waves, the two-dimensional power spectral density, $S_{k\omega} = \log_{10}|\hat{\eta}_d(k,\omega_r)|^2$, is calculated from the interface elevation data, $\eta(x,t)$ in the nonlinear state. Instantaneous spatial mean is subtracted from the data, and a one-dimensional Hann window \citep{oppenheim1989discrete}, $w(t)$, is applied along the time axis:$$\tilde{\eta}(x,t) = \left[ \eta(x,t) - \langle \eta(x,t) \rangle_x \right] w(t).$$The windowed signal is transformed into the spectral domain using a two-dimensional Fast Fourier Transform (FFT). The log-scaled PSD is normalized by the domain size. To isolate the peak spectral components, we employ the \texttt{imregionalmax} function in MATLAB \citep{MATLAB:R2025a}.

\section{$H_1,H_2$ reconstruction}\label{AppC:H2_H2_recon}
Here we reconstruct the interfaces $H_1,H_2$ from the respective peak spectral modes listed in tables \ref{tab:Tab4_bound_harmonics_H1} and \ref{tab:modes_H2} for $Fr = 0.8$.
\begin{figure}
	\centerline{\includegraphics[width = 0.8\textwidth]{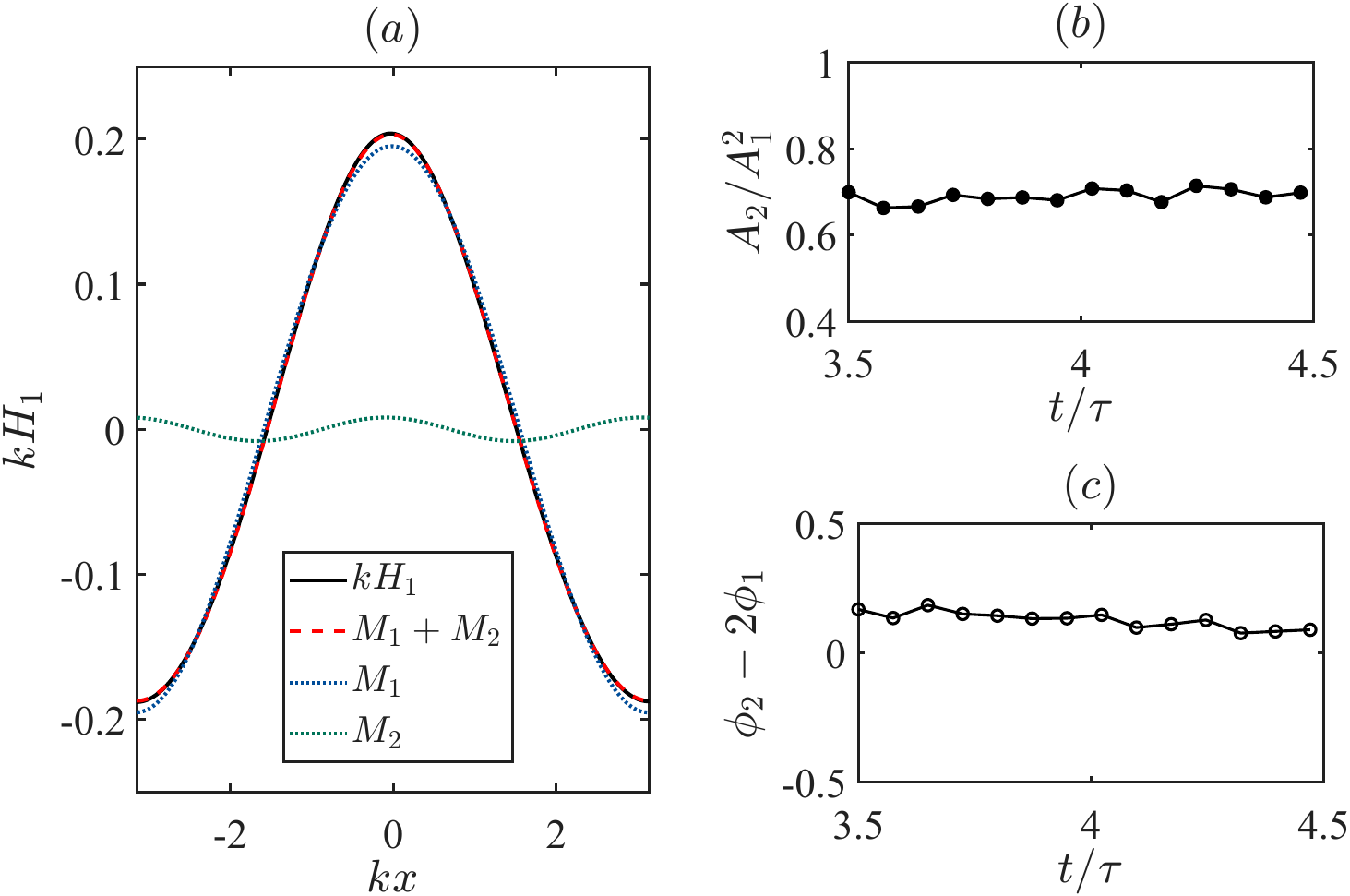}}
	\caption{Reconstruction of the density-vorticity interface $H_1$ using the dominant spectral components extracted from FFT for $Fr = 0.8$ (component properties detailed in Table~\ref{tab:Tab4_bound_harmonics_H1}). Individual modes are defined as $M_n = A_n \cos(k_n x + \phi_n)$, where $k_n = n k_0$. (a) Snapshot of the non-dimensionalized density interface $kH_1$ versus $kx$ at time $t/\tau = 4$. The raw interface from the simulation (solid black curve) is compared against the reconstructed profile $M_1 + M_2$ (dashed red curve). The individual carrier mode $M_1$ (dotted blue curve) and second bound harmonic $M_2$ (dotted green curve) are also shown. The near-perfect overlap indicates that the first two components sufficiently capture the nonlinear wave shape. (b) Temporal evolution of the amplitude ratio $A_2/A_1^2$ and (c) the phase difference $\phi_2 - 2\phi_1$ over the time window $3.5 \le t/\tau \le 4.5$. The near constancy of these values confirms that $M_2$ acts as a bound harmonic phase-locked to the primary carrier wave. Fixed physical parameters are identical to those in Figure~\ref{fig9:LSA_basilisk_H1}.}
	\label{fig:H1_reconstruction}
\end{figure}
\begin{figure}
	\centerline{\includegraphics[width = 0.8\textwidth]{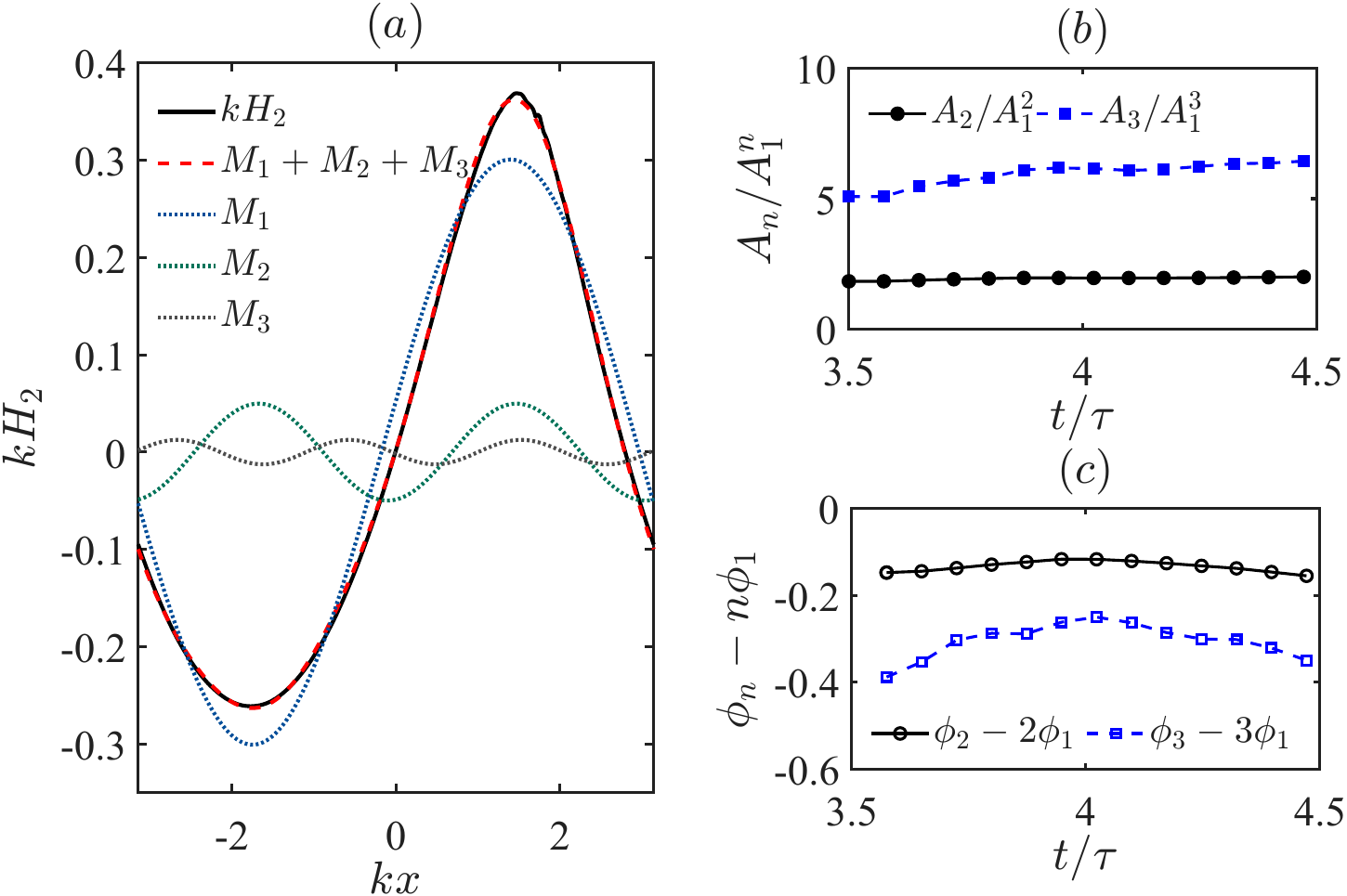}}
	\caption{Reconstruction of the subsurface vorticity interface $H_2$ using the first three dominant spectral components ($M_1, M_2, M_3$) for $Fr = 0.8$ (see Table \ref{tab:modes_H2} for mode properties). (a) Snapshot of the non-dimensionalized vorticity interface $kH_2$ versus phase $kx$ at $t/\tau = 4$. The raw simulation interface (solid black curve) is compared against the reconstructed profile $M_1 + M_2 + M_3$ (dashed red curve). Because the vorticity interface is steeper than the density interface, three components are required to accurately capture its shape. (b) Temporal evolution of the amplitude ratios $A_2/A_1^2$ and $A_3/A_1^3$. (c) Evolution of the modified phase differences $\phi_2 - 2\phi_1$ and $\phi_3 - 3\phi_1$. The nearly constant amplitude ratios and phase differences over $3.5 \le t/\tau \le 4.5$ confirm that $M_2$ and $M_3$ are bound harmonics phase-locked to the carrier wave.}
	\label{fig:H2_reconstruction}
\end{figure}
Figs. \ref{fig:H1_reconstruction} and \ref{fig:H2_reconstruction} visually confirm the accuracy of the modal reconstruction. Fig. \ref{fig:H1_reconstruction}  Panel (a) demonstrates that superimposing just the fundamental carrier mode ($M_1$) and the second bound harmonic ($M_2$) yields a near-perfect match with the raw simulated density interface ($kH_1$). Panels (b) and (c) further validate the physical nature of these components: both the amplitude ratio ($A_2/A_1^2$) and phase difference ($\phi_2 - 2\phi_1$) remain remarkably constant over the observed time window. This is a classical signature of steady nonlinear sheared Stokes wave, confirming that $M_2$ is phase-locked to the primary carrier mode as a bound harmonic.
In fig. \ref{fig:H2_reconstruction}, since the vorticity wave is steeper (i.e. more nonlinear) than the density-vorticity wave, three peak spectral components ($M_1$, $M_2$, and $M_3$) are required to accurately capture its profile. Panel (a) demonstrates that superimposing these three modes yields an excellent match to the raw simulated interface ($kH_2$). Panels (b) and (c) further validate the physical nature of these higher-order components. Over the observed time window, both the amplitude ratios ($A_n/A_1^n$) and the phase relationships ($\phi_n - n\phi_1$) remain almost constant. This temporal stability confirms that $M_2$ and $M_3$ are phase-locked to the primary carrier mode as bound harmonics.
 
\section{Nonlinear steady state solution details and validation}\label{AppD:NonlinearBVP}
\subsection{Algebraic residual equations for the fully nonlinear steady state}
\label{app:residual_equations}

To compute the finite-amplitude permanent-form solutions, the continuous boundary conditions are discretized at $2N+1$ equally spaced collocation points, defined as $\tilde{x}_i = 2\pi(i-1)/(2N+1)$ for $i = 1, 2, \dots, 2N+1$. At each discrete point, the local interfacial elevations and their analytical spatial derivatives are evaluated directly from their Fourier expansions:
\begin{equation*}
    \tilde{H}_{1,i} = \sum_{n=1}^N \Big[\alpha_n \cos(n\tilde{x}_i) + \hat{\alpha}_n \sin(n\tilde{x}_i)\Big], \quad \tilde{H}_{2,i} = -\tilde{\Delta} + \sum_{n=1}^N \Big[d_n \cos(n\tilde{x}_i) + \hat{d}_n \sin(n\tilde{x}_i)\Big],
\end{equation*}
\begin{equation*}
    \tilde{H}_{1,i}' = \sum_{n=1}^N n \Big[-\alpha_n \sin(n\tilde{x}_i) + \hat{\alpha}_n \cos(n\tilde{x}_i)\Big], \quad \tilde{H}_{1,i}'' = -\sum_{n=1}^N n^2 \Big[\alpha_n \cos(n\tilde{x}_i) + \hat{\alpha}_n \sin(n\tilde{x}_i)\Big].
\end{equation*}
Substituting the truncated Fourier series \eqref{eq: nondim_fourier_series} into the boundary conditions \eqref{eq: nondim_BC_FA}-\eqref{eq: BC_norm_stress_FA_nondim} yields the following system of algebraic residual equations, applied simultaneously at every collocation point $\tilde{x}_i$:
\begin{subequations}
    \begin{align}
       & (\tU_s - \tilde{c})\tilde{H}_{1,i} + Q_a + \sum_{n=1}^N e^{-n\tilde{H}_{1,i}}\Big[g_n \cos{(n\tilde{x}_i)}+ \hat{g}_n \sin{(n\tilde{x}_i)}\Big] = 0, \\[8pt]
       %
       & \frac{\tilde{\Omega}\tilde{H}_{1,i}^{2}}{2} - \hat{\tilde{c}}\tilde{H}_{1,i} + Q_{w_1} + \sum_{n=1}^N \Big[\cos{(n\tilde{x}_i)}\big\{a_n e^{n\tilde{H}_{1,i}} + b_n e^{-n (\tilde{H}_{1,i}+\tilde{\Delta})}\big\} \nonumber \\
       & \quad + \sin{(n\tilde{x}_i)}\big\{\hat{a}_n e^{n \tilde{H}_{1,i}} + \hat{b}_n e^{-n(\tilde{H}_{1,i}+\tilde{\Delta})}\big\}\Big]= 0, \\[8pt]
       %
       & \frac{\tilde{\Omega}\tilde{H}_{2,i}^{2}}{2} - \hat{\tilde{c}}\tilde{H}_{2,i} + Q_{w_1} + \sum_{n=1}^N \Big[\cos{(n\tilde{x}_i)}\big\{a_n e^{n \tilde{H}_{2,i}} + b_n e^{-n (\tilde{H}_{2,i}+\tilde{\Delta})}\big\} \nonumber \\
       & \quad + \sin{(n\tilde{x}_i)}\big\{\hat{a}_n e^{n \tilde{H}_{2,i}} + \hat{b}_n e^{-n (\tilde{H}_{2,i}+\tilde{\Delta})}\big\}\Big]= K_1, \\[8pt]
       %
       & \frac{\tilde{\Omega}\tilde{H}_{2,i}^{2}}{2} + (\tilde{c}-\hat{\tilde{c}})\tilde{H}_{2,i} + Q_{w_1} - Q_{w_2} + \sum_{n=1}^N \Big[\cos{(n\tilde{x}_i)}\big\{a_n e^{n \tilde{H}_{2,i}} + b_n e^{-n (\tilde{H}_{2,i}+\tilde{\Delta})} \nonumber \\ 
       & \quad - f_n e^{n (\tilde{H}_{2,i}+\tilde{\Delta})}\big\} + \sin{(n\tilde{x}_i)}\big\{\hat{a}_n e^{n \tilde{H}_{2,i}} + \hat{b}_n e^{-n (\tilde{H}_{2,i}+\tilde{\Delta})} - \hat{f}_n e^{n (\tilde{H}_{2,i}+\tilde{\Delta})}\big\}\Big] = 0, \\[8pt]
       %
       & \tilde{\Omega}\tilde{H}_{2,i} + \tilde{c} - \hat{\tilde{c}} + \sum_{n=1}^N n\Big[\cos{(n\tilde{x}_i)}\big\{a_n e^{n \tilde{H}_{2,i}} - b_n e^{-n (\tilde{H}_{2,i}+\tilde{\Delta})} - f_n e^{n (\tilde{H}_{2,i}+\tilde{\Delta})}\big\} \nonumber \\
       & \quad + \sin{(n\tilde{x}_i)}\big\{\hat{a}_n e^{n \tilde{H}_{2,i}} - \hat{b}_n e^{-n (\tilde{H}_{2,i}+\tilde{\Delta})} - \hat{f}_n e^{n (\tilde{H}_{2,i}+\tilde{\Delta})}\big\}\Big] = 0, \\[8pt]
       %
       & \frac{1}{2}\Bigg(\sum_{n=1}^N n\Big[-\sin{(n\tilde{x}_i)}\big\{a_n e^{n \tilde{H}_{1,i}} + b_n e^{-n (\tilde{H}_{1,i}+\tilde{\Delta})}\big\} + \cos{(n\tilde{x}_i)}\big\{\hat{a}_n e^{n \tilde{H}_{1,i}} + \hat{b}_n e^{-n (\tilde{H}_{1,i}+\tilde{\Delta})}\big\}\Big] \Bigg)^2 \nonumber \\
       & \quad + \frac{1}{2}\Bigg(\tilde{\Omega}\tilde{H}_{1,i} - \hat{\tilde{c}} + \sum_{n=1}^N n\Big[\cos{(n\tilde{x}_i)}\big\{a_n e^{n \tilde{H}_{1,i}} - b_n e^{-n (\tilde{H}_{1,i}+\tilde{\Delta})}\big\} \nonumber \\
       & \quad + \sin{(n\tilde{x}_i)}\big\{\hat{a}_n e^{n \tilde{H}_{1,i}} - \hat{b}_n e^{-n (\tilde{H}_{1,i}+\tilde{\Delta})}\big\}\Big] \Bigg)^2 - \frac{r}{2}\Bigg(\sum_{n=1}^N n\:e^{-n\tilde{H}_{1,i}}\Big[-g_n \sin{(n\tilde{x}_i)} + \hat{g}_n \cos{(n\tilde{x}_i)}\Big]\Bigg)^2 \nonumber \\ 
       & \quad - \frac{r}{2} \Bigg(\tU_s - \tilde{c} - \sum_{n=1}^N n\:e^{-n\tilde{H}_{1,i}}\Big[g_n \cos{(n\tilde{x}_i)} + \hat{g}_n \sin{(n\tilde{x}_i)}\Big]\Bigg)^2 \nonumber \\
       & \quad + \tilde{H}_{1,i}(1-r) - \frac{\tT\tilde{H}''_{1,i}}{\big(1+[\tilde{H}'_{1,i}]^2\big)^{3/2}} - B_1 = 0.
    \end{align}
    \label{eq:nonlinear_system_eqs}
\end{subequations}
Finally, we have additional global wave height constraint \eqref{eq: wave height} and the condition \eqref{eq: phase_pinning} at $\tilde{x}_1 = 0$, both of which are discussed in \S \ref{sec:Connection to nonlinear waves}.

\subsection{Initial guess from linear stability analysis}\label{app:initial_guess_LSA}

To initiate the numerical continuation in $h$ (from \S \ref{subsec:continuation}), the analytically estimated eigenmodes as derived from the linear stability analysis (\S \ref{sec:Linear_stability}) are utilized as the initial guess. This requires mapping the time-dependent, linearly scaled perturbations onto the steady, wave-scaled Fourier series. 

We demonstrate this projection explicitly for the upper density-vorticity interface. In the linear stability framework, the dimensional interfacial perturbation evaluated at $t^*=0$ is given by $H_1^*(x^*) = \text{Re}\{\eta^*_d\: e^{i k^*x^*}\}$. Without loss of generality, we fix the spatial phase of the wave by taking the complex amplitude $\eta^*_d$ to be strictly real, yielding $H_1^*(x^*) = \eta^*_d \cos(k^* x^*)$. In the steady finite-amplitude formulation \eqref{eq: nondim_fourier_series}(\textit{a}), the fundamental harmonic of this interface is represented as $\tilde{H}_1(\tilde{x}) = \alpha_1 \cos(\tilde{x}) + \hat{\alpha}_1 \sin(\tilde{x})$. By equating these functional forms and accounting for the transition in characteristic length scales, from the shear depth $\Delta^*$ in the linear analysis to the inverse wavenumber $1/k^*$ in the nonlinear analysis, we obtain the exact mapping for the leading-order coefficients:
\begin{equation}
    \alpha_1 = k^* \eta_d^* = (k^*\Delta^*) \left(\frac{\eta_d^*}{\Delta^*}\right) = k \eta_d, \quad \text{and} \quad \hat{\alpha}_1 = 0,
    \label{eq:H1_initial_guess}
\end{equation}
where $k$ and $\eta_d$ are the non-dimensional wavenumber and interfacial amplitude extracted from the linear stability solution. 

Applying this identical phase-locking and scale conversion to the subsurface vorticity interface and the layer-specific stream functions yields the complete set of initial non-zero fundamental Fourier coefficients required to seed the solver. Employing the complex linear stability variables from \eqref{eq: LSA_coefficients}, the remaining Fourier coefficients for the nonlinear boundary value problem are mapped as follows:
\begin{subequations}
    \begin{align}
        d_1 &= k\,\text{Re}(\eta_v), &&\hat{d}_1 = -k\,\text{Im}(\eta_v), \\[4pt]
        g_1 &= k \sqrt{k}\, Fr\, \text{Re}(A), &&\hat{g}_1 = -k \sqrt{k}\, Fr\, \text{Im}(A), \\[4pt]
        a_1 &= k \sqrt{k}\, Fr\,\text{Re}(B), &&\hat{a}_1 = -k \sqrt{k}\, Fr\,\text{Im}(B), \\
        b_1 &= k \sqrt{k}\, Fr\,\text{Re}(D)\,e^{k}, &&\hat{b}_1 = -k \sqrt{k}\, Fr\,\text{Im}(D)\,e^{k}, \\[4pt]
        f_1 &= k \sqrt{k}\, Fr\,\text{Re}(E)\,e^{-k}, &&\hat{f}_1 = -k \sqrt{k}\, Fr\,\text{Im}(E)\,e^{-k}.
    \end{align}
    \label{eq:Fourier_mappings}
\end{subequations}
The non-dimensional phase speeds and the global wave height parameter are initialized as:
\begin{equation}
    \tilde{c} = Fr \sqrt{k}\,\text{Re}(c), \quad \hat{\tilde{c}} = Fr \sqrt{k}\,(\text{Re}(c) - 1), \quad \text{and} \quad h = 2\alpha_1 = 2 k \eta_d.
\end{equation}

Finally, to ensure the iterative solver initializes within the correct basin of attraction, the integration constants ($Q_a, Q_{w_1}, Q_{w_2}, K_1, B_1$) must also be initialized. These are determined analytically by evaluating the residual equations \eqref{eq:nonlinear_system_eqs} at the unperturbed trivial base state ($\tilde{H}_1 = 0$, $\tilde{H}_2 = -\tilde{\Delta}$, with all Fourier wave coefficients set to zero), which yields:
\begin{subequations}
    \begin{align}
        Q_a &= 0, \\
        Q_{w_1} &= 0, \\
        Q_{w_2} &= \frac{1}{2}\tilde{\Omega}\tilde{\Delta}^2 - (\tilde{c} - \hat{\tilde{c}})\tilde{\Delta}, \\
        K_1 &= \frac{1}{2}\tilde{\Omega}\tilde{\Delta}^2 + \hat{\tilde{c}}\tilde{\Delta}, \\
        B_1 &= \frac{1}{2}\hat{\tilde{c}}^2 - \frac{r}{2}(\tU_s - \tilde{c})^2.
    \end{align}
\end{subequations}
Together with the linear modes mapped in \eqref{eq:Fourier_mappings}, these constants provide the initial guess required to initiate numerical continuation in wave height $h$.

\subsection{Validation of the nonlinear finite-amplitude analysis}\label{appC:validation_FA}
\begin{figure}[ht!]
  \centering
  \includegraphics[width = \textwidth]{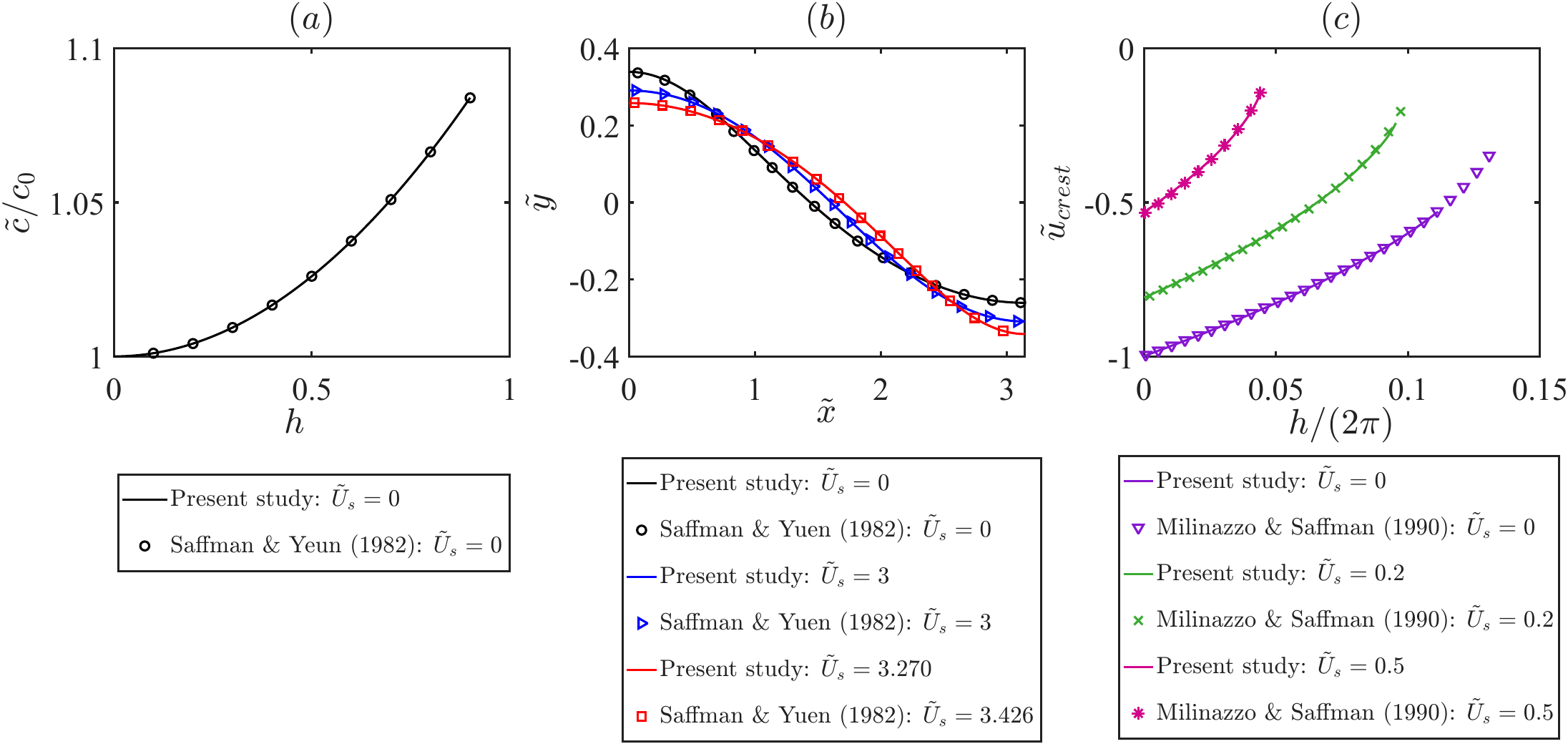}
  \caption{Validation of the algorithm used for computing nonlinear progressive waves of permanent form in \S \ref{subsec:Nonlinear_formulation}, \S\ref{subsec:continuation} against literature. (a) Normalized wave phase speed, $\tc/c_0$, as a function of the density-vorticity wave height $h$, where $c_0 = \sqrt{(1-r)/(1+r)}$ represents the linearised limit of the phase-speed. (b) Permanent interfacial wave profiles at $h = 0.6$ for varying free-stream air velocities $\tilde{U}_s$. For both (a) and (b), solid lines denote the present formulation while symbols represent the data of \citet{saffman1982finite}, evaluated for an irrotational water layer ($\tilde{\Omega} = 0$), zero surface tension ($\tilde{T} = 0$), and a density ratio of $r = 0.1$. (c) Horizontal fluid velocity at the wave crest, $\tilde{u}_{crest}$, as a function of the normalized wave height $h/(2\pi)$ for different values of surface drift ($\tilde{\Omega}\tilde{\Delta} = \tU_s$). Solid lines denote the present formulation and symbols represent the results of \citet{milinazzo1990effect}, evaluated with a constant shear depth of $\tilde{\Delta} = 0.05$ and $\tilde{T} = 0$.} 
\label{fig:validation_saffman}
\end{figure}

To validate the numerical framework in the limit of irrotational flow, we first compare our results against the study of \citet{saffman1982finite}, who computed permanent-form solutions for finite-amplitude interfacial gravity waves ($\tilde{T} = 0$) subjected to a uniform air current ($\tU_s$). Figure \ref{fig:validation_saffman}(\textit{a}) plots the normalized phase speed, $\tilde{c}/c_0$ (where $c_0 = \sqrt{(1-r)/(1+r)}$), as a function of the wave height $h$. For an unforced, irrotational system ($\tU_s = 0$, $\tilde{\Omega} = 0$) at a density ratio of $r = 0.1$, our solver demonstrates very good quantitative agreement with their data. Furthermore, figure \ref{fig:validation_saffman}(\textit{b}) compares the spatial wave profiles evaluated at a highly nonlinear wave height of $h = 0.6$ across various free-stream wind speeds. Our computed finite-amplitude shapes (solid lines) perfectly trace the profiles reported by \citet{saffman1982finite} (symbols).

To validate the solver's capability to capture strongly rotational kinematics, we compare our formulation against the results of \citet{milinazzo1990effect}. They investigated steady finite-amplitude waves in a deep-water system without an overlying air layer ($r = 0$), where the wind drift was modeled by a thin, uniform subsurface vorticity layer, analogous to the rotational base state of the present study. Figure \ref{fig:validation_saffman}(\textit{c}) displays the horizontal fluid velocity at the wave crest, $\tilde{u}_{crest}$, as a function of the normalized wave steepness, $h/(2\pi)$. The profiles are evaluated across multiple surface drift velocities ($\tilde{\Omega}\tilde{\Delta} = \tU_s$) for a fixed shear depth of $\tilde{\Delta} = 0.05$ and zero surface tension ($\tilde{T} = 0$). Our computed solutions (solid lines) exhibit good quantitative agreement with their benchmark data (symbols). Importantly, the permanent-form solutions reported by these authors correspond to stable flat base states, wherein the density and vorticity interfaces remain symmetric.

\bibliographystyle{jfm}
\bibliography{jfm}

\end{document}